\documentclass[a4paper]{article}

\usepackage{graphicx}
\usepackage[numbers]{natbib}

\usepackage{upgreek} 
\usepackage{amsmath,amssymb,bm} 

\usepackage{pgffor}
\usepackage{tikz}
\usetikzlibrary{shapes,arrows,fit}
\usetikzlibrary {calc,math,arrows.meta}
\usetikzlibrary {decorations.pathmorphing}
\tikzstyle{every loop}=[]
\tikzset{
physfront/.style = {
         rectangle, draw,
         minimum width=1.7cm, minimum height=1cm, 
         color=orange, text=black, fill=orange!25}
}
\tikzset{
physback/.style = {
         rectangle, 
         minimum width=1.7cm, minimum height=1cm, 
         text=black, fill=orange!15}
}
\tikzset{
absfront/.style = {
         rounded rectangle, draw,
         minimum width=2cm, minimum height=1cm, 
        color=blue, text=black, fill=blue!20} 
}
\tikzset{
absback/.style = {
         rounded rectangle, 
         minimum width=2cm, minimum height=1cm, 
        text=black, fill=blue!10} 
}
\tikzset{
physphysfront/.style = {
         -{Triangle[open]}, 
         color=orange,  thick,
         text = black} 
}
\tikzset{
physphysback/.style = {
         -{Triangle[open]}, 
         color=orange, very thick, dashed,
         text = black} 
}
\tikzset{
physphysside/.style = {->, font=\normalsize,align = center} 
}
\tikzset{
absabsfront/.style = {
         -{Triangle[open]}, 
         color=blue,  thick,
         text = black} 
}
\tikzset{
absabsback/.style = {
         -{Triangle[open]}, 
         color=blue, very thick, dashed,
         text = black} 
}
\tikzset{
absabsside/.style = {->, font=\normalsize,align = center} 
}

\newcommand{\bigp}{\mathbf{P}}
\newcommand{\p}{\mathbf{p}}

\newcommand{\RT}{\mathcal{R_T}}
\newcommand{\RRE}{\mathcal{R}_{RE}}
\newcommand{\Rc}{\mathcal{R}_{c}}
\newcommand{\RTRE}{\mathcal{R}_{\mathcal{T}_{RE}}}
\newcommand{\IRT}{\mathcal{\widetilde{R}_T}}
\newcommand{\IRc}{\mathcal{\widetilde{R}}_{c}}
\newcommand{\IRRE}{\mathcal{\widetilde{R}}_{RE}}
\newcommand{\IRTc}{\mathcal{\widetilde{R}}_{\mathcal{T}_{c}}}
\newcommand{\theory}{\mathcal{T}}
\newcommand{\see}{C}

\newcounter{notebox}
\usepackage[colorinlistoftodos]{todonotes}

\begin{document}

\title{When does a control system compute?\\
Digital, mechanical and open-loop systems
}

\author{
Dominic Horsman$^{1}$, 
Susan Stepney$^{1}$, 
Tim~Clarke$^{2}$,
and Viv Kendon$^{3}$
}

\date{\small
$^{1}$Department of Computer Science, University of York, UK
\\
\texttt{susan.stepney@york.ac.uk}
\\
$^{2}$School of Physics, Engineering and Technology, University of York, UK 
\\
$^{3}$Department of Physics, University of Strathclyde, Glasgow, UK
\\
\texttt{viv.kendon@strath.ac.uk}
}

\maketitle

\begin{abstract}
Control systems are ubiquitous in modern technology, comprising an engineered \textit{plant} to be kept within specific, often fine-tuned, limits, and a separate \textit{controller} that ensures this is the case. 
While modern controllers often employ digital computers, other examples are purely mechanical, or even biological. It is an open question whether computation is happening within all controllers by virtue of them being part of a control system. 
Abstraction/\hspace{0pt}Representation theory (ART) has been developed to tackle just this question of whether a physical system is computing. 
Here, we  demonstrate how to use ART to model control systems, and analyse them for computational properties. We determine that the plant of a control system is (a proxy for) the representational entity necessary in ART for the existence of any computation: the plant is the \textit{user} of the controller. 
We consider specific systems: a digital thermostat, an electro-mechanical thermostat, the purely mechanical centrifugal governor, and an open-loop human-controlled heating system. 
We show that all these systems, and control systems in general, are performing some degree of computation. 
As an initial use of these results, we apply them to
computationalism within cognitive theory: we show the governor \textit{is} computing, so it cannot  play its role of counter-example in the question of whether the brain is too.
\end{abstract}

\section{Introduction}

Here we explore the nature of control systems and their relationship to the process of computing.
Control systems are often complex, interconnected, dynamical systems with feedback. 
They are widespread in engineering, including mechanical and electronic applications, chemical processing, and manufacturing.
Examples include robotic arm control systems in manufacturing processes, feedback control systems in power grid management, and guidance and navigation systems in spacecraft.
In all cases, the dynamics of one system (the \textit{plant}) are controlled, or regulated, by a second system (the \textit{controller}), based on external variables, or on properties of the plant, or a combination of the two.
The many key applications of control systems make it important to understand their behaviour, and what they are capable of, in computational terms.

At first glance, control systems bear many resemblances to systems that are computing. The adjustment of behaviour based on other system variables appears similar to the processing and passing of information through a computing system. 
Indeed, many modern control systems incorporate digital computers to perform some of the tasks within the control system loop. The question we consider here is whether this is more than a mere resemblance: does a control system (any or all of them) \textit{compute}?

The apparent similarities between control systems and computing systems has been discussed by various authors, coming to different apparent conclusions. 
The mathematics describing control systems is, via the underlying category theory, almost identical with the mathematics of the ZX-calculus used for quantum computation \cite{Baez-Erbele}; 
control systems signal flow diagrams have a symmetric monoidal structure \cite{Fong-Pawel}, again like that for some formulations of quantum computation. 
Contrastingly, one particular control system has been used as an example of a device that is \textit{not} computing. The centrifugal governor is an Industrial Revolution-era device used to regulate the amount of steam in an engine using a pair of flyballs to operate a valve controlling steam pressure. This is a purely mechanical system, and perhaps easier to see as a `pure' control system than, for example, a nuclear power plant with digital computers as part of the control process. This system has been used as an example to argue against the notion that cognition must be computation  \cite{Van_Gelder1995,Baltieri2020}, by arguing there are complex dynamical systems that look like computation but are not. The centrifugal governor control system is considered a dynamical system that is not computing.
Here we examine the performance of generic control systems, and come to a different conclusion.

In section~\ref{sec:art} we summarise our approach to defining physical computing through Abstraction/Representation theory (ART).
In section~\ref{sec:ctrlsys} we show how control systems can be modelled through the lens of ART,
by translating control system diagrams into ART diagrams.
In section~\ref{sec:control_ART} we provide examples of digital, electro-mechanical, mechanical, and human-in-the-loop control systems, including the centrifugal governor,
and show in detail how and where the computing is happening.
We conclude that \textit{all control systems, whether digital electronic, purely mechanical, human-in-the-loop, or otherwise, are computing}.
In section~\ref{sec:discussion} we discuss several implications of our analysis.
These include the (mis)use of the centrifugal governor as a counterexample to computation in cognitive theory,
and suggest that discipline should choose a different counterexample;
and the identification of the deep connection, and fundamental differences, between computing and control systems.

\section{Our view of physical computing: ART in a nutshell}\label{sec:art}
\subsection{Why ART}

Understanding when any physical system, control or otherwise, can be said to be computing is a highly non-trivial question. Many attempts have been made in the last century to tackle this issue, and the answer to this general question is key for our specific investigation into whether control systems can be classified as computing.

The point of view that everything computes all the time, and that every physical system is a computer simply by virtue of being a physical system \cite{Lloyd07}, can immediately be set aside. It is either trivially true, and hence uninformative, or else false. In neither case does it allow us to differentiate instances of computing from non-computing.
Other views that equate computing with the process that can be performed on Turing Machines \citep{turing37} are also not useful for our current purpose. To begin with, they ignore the existence of computation outside this paradigm, such as analogue computing. Analogue computing devices, such as the differential analyser, were common in the early 20th century. Their computational abilities were formalised by Shannon in his GPAC (General Purpose Analogue Computing) model \citep{Shannon41}, which is not interchangeable with the Turing Machine model. 
Moreover, equating computing with Turing Machine-capable computation ignores the physical status of the computer itself. Turing computation concerns the functioning of a purely abstract device, with no physical substrate; it is an abstract, mathematical, concept of \textit{computation}. By contrast, the notion of \textit{computing} includes both the abstract computation and the physical computing device on which it is supported.

We use here the framework of Abstraction/Representation Theory (ART), which was developed (including by three of the present authors) to finally answer this question of when a physical system computes \citep{Horsman2014}. It is a diagrammatic pseudo-formal framework that represents computing along with science, engineering/technology, and communication, all using the same basic elements of \textit{representational activity}. Computing is a precisely-specified process with elements that must be present, and conditions that must be met, in order for it to be said to be occurring.

ART provides a framework in which science, engineering/technology, computing, and communication/signalling are all defined as \textit{representational activity} requiring the fundamental use of the representation relation in order to define their operation \citep{Horsman2014,Horsman2015}.
In work following on from the original definitions,
Horsman et al.
provide a high level overview \cite{Horsman2017},
delve into more philosophical aspects \cite{Horsman2018},
and present an example of intrinsic computation : signalling in bacteria, showing that the bacterium is its own representational entity \cite{Horsman2017bio}.
In order to discuss control systems, the role of the representational entity during a compute-cycle has been  formally identified
\cite{StepneyKendon2019}.

ART has been developed to answer the specific question of when a physical system is computing \citep{Horsman2014}. The answer hinges on the relationship between an abstract object (a computation) and a physical object (a computer). It employs a language of relations, not from mathematical objects to mathematical objects (as is usual in mathematics and theoretical computer science), but between physical objects and those in the abstract domain. 
The core of ART is the \textit{representation relation}, mapping from physical objects to abstract objects. Experimental science, engineering, and computing all require the interplay of abstract and physical objects via representation in such a way that their descriptive diagrams commute: the same result can be gained through either physical or abstract evolutions.
From this comes the  definition of computing as \textit{the use of a physical system to predict the outcome of an abstract evolution}. 

\subsection{Representation}

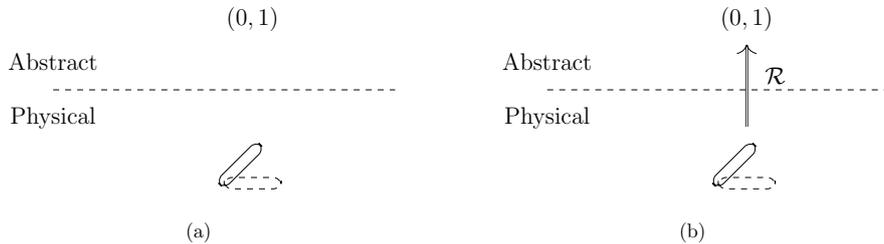
\begin{figure}[t]
    \scalebox{0.75}{

\begin{minipage}[c]{1.0\linewidth}
  \[
   \begin{array}{ccc}
   \begin{tikzpicture}[font=\large]

\draw[style=dashed] (0,0) -- (6,0);
\draw (0,0.5) node {Abstract};
\draw (0,-0.5) node {Physical};

\draw[rounded corners,rotate around={45:(3,-1.75)}] (3,-1.75) rectangle (4,-1.55);
\draw[rounded corners,dashed] (3,-1.75) rectangle (4,-1.55);

\draw (3.5,1.25) node {$(0,1)$};

\end{tikzpicture}
& \qquad 
\quad &
\begin{tikzpicture}[font=\large]

\draw[style=dashed] (0,0) -- (6,0);
\draw (0,0.5) node {Abstract};
\draw (0,-0.5) node {Physical};

\draw[rounded corners,rotate around={45:(3,-1.75)}] (3,-1.75) rectangle (4,-1.55);
\draw[rounded corners,dashed] (3,-1.75) rectangle (4,-1.55);

\draw (3.5,1.25) node {$(0,1)$};

\draw[->,double] (3.5,-0.65) -- (3.5,.8);
\draw (4,0.25) node {$\mathcal{R}$};

\end{tikzpicture}
\\{}\\
 \mathrm{(a)} & \qquad \quad  &  \mathrm{(b)}  \end{array}
\]
\end{minipage}}
\caption{\label{reprel}Basic representation. (a) Spaces of abstract and physical objects (here, a switch with two settings and a binary digit). (b) The directed representation relation $\mathcal{R}$ mediating between the spaces.} 
\end{figure}

ART contains \textit{physical objects} in the domain of material systems, \textit{abstract objects} (including mathematical and logical entities), and the \textit{representation relation} that mediates between the two. The distinction between the two spaces, abstract and physical, is fundamental in the theory, as is their connection \textit{only} by the (directed) representation relation. An intuitive example is given in figure~\ref{reprel}: a physical switch is \textit{represented} by an abstract bit, where in this case the bit takes the value 0 for switch state up, and 1 for switch state down.
It would be equally valid to take the opposite mapping of 1 for up and 0 for down:
representation involves arbitrary choices.

The central role of representation leads to the requirement for a \textit{representational entity} (RE). 
The RE supports the representation relation between physical and abstract. 
ART does not require the RE to be human, or  conscious; see \cite{Horsman2017bio,StepneyKendon2019} for an example of a bacterial RE.

The elementary \textit{representation relation} is the directed map from physical to abstract objects, \mbox{$\mathcal{R_T} : \bigp \rightarrow M$},
where $\bigp$ is the set of physical objects, and $M$ is the set of abstract objects. 
We subscript the relation $\mathcal{R}$ with a theory $\mathcal{T}$ to indicate that the relation is theory-dependent.
When a physical object $\p$ and an abstract object $m_\p$ are connected by $\mathcal{R_T}$ we write them as $\p \mapsto m_\p$. The abstract object is then said to be the \textit{abstract representation} (under the given theory) of the physical object.
This basic representation is shown in figure \ref{physabst}.

Abstract evolution takes abstract objects to abstract objects, which we write as $C_\mathcal{T}: M \rightarrow M$. 
Again, we subscript with theory $\mathcal{T}$ to indicate that $C$ is theory-dependent.
An individual example is shown in figure \ref{physabst}, for the mapping $C_\mathcal{T}(m_\p)$ taking $m_\p \mapsto m_\p$. The corresponding physical evolution map is given by \mbox{$\mathbf{H}: \bigp \rightarrow \bigp$}. For individual elements in figure \ref{physabst}(c) this is $\mathbf{H(p)}$ which takes $\p \mapsto \p$.

\subsection{$\epsilon$-commuting diagrams}

\begin{figure}[t]
\centering
    \scalebox{0.7}{\begin{tikzpicture}[font=\large]

\draw[style=loosely dashed] (1,0) 
    -- node[above,at start] {\small{\textit{Abstract}}}
     node[below,at start] {\small{\textit{Physical}}}
    (9.5,0);

\node[physfront] (p) at (3,-1) {$\strut\p$};
\node[absfront] (mp) at (3,2) {$\strut m_{\p}$};
\draw[->,double] (p) --  node[right] {$\RT$} (mp);

\node[absfront] (mprp) at (6.5,2) {$\strut m^\prime_{\p}$};
\draw[absabsfront] (mp) -- node[above] {$C_\theory(m_\p)$} (mprp);

\node[physfront] (ppr) at (8,-1) {$\strut\p^\prime$};
\draw[physphysfront] (p) -- node[above] {$\mathbf{H}(\p)$} (ppr);

\node[absfront] (mppr) at (8,1) {$\strut m_{\mathbf{p}^\prime}$};
\draw[->,double] (ppr) -- node[right,near start] {$\RT$} (mppr);

\draw [<->,shorten <=.1cm,shorten >=.1cm] (mppr) to [out=90,in=0] (mprp);
\draw (8.1,1.9) node {$\varepsilon$};

\end{tikzpicture}}
\caption{Parallel evolution of an abstract object (blue, round corners) and the physical system (orange, square corners) it represents. 
Left line: the basic representation: physical system $\p$ is represented abstractly by $m_\p$ using the modelling representation relation $\RT$ of theory $\theory$. 
Top line: abstract dynamics $\see_\theory (m_\p)$ give the evolved abstract state $m^\prime_\p $. 
Bottom line: physical dynamics $\mathbf{H}(\p)$ give the final physical state $\p^\prime$. 
Right line: $\RT$ is used again to represent $\p^\prime$ as the abstract output $m_{\p^\prime}$, $|m_\p - m_{\p^\prime}| = \varepsilon$. 
(Adapted from \cite{Horsman2014}.)}\label{physabst}
\end{figure}
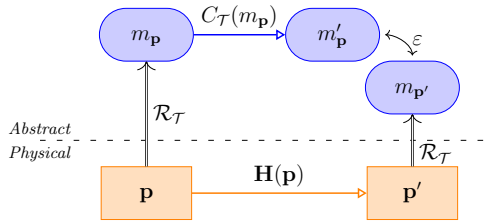

In figure~\ref{physabst}, to link the final abstract and physical objects, we apply the representation relation to the outcome state of the physical evolution, to give its abstract representation $m_{\p}$. We now have two abstract objects: $m_\p$ is the result of the abstract evolution; $m_{\p}$ is the representation of the result of the physical evolution. 
For some (problem-dependent) error quantity $\varepsilon$ and distance function $d()$, if $d(m_{\p}, m_{\p}) \le \varepsilon$ (or, more briefly, $m_{\p} =_\varepsilon m_{\p}$), then we say that the diagram in figure \ref{physabst} $\varepsilon$-\textit{commutes}. 
Commuting diagrams are fundamental to the use of ART. If a set of abstract and physical objects form an $\varepsilon$-commuting diagram under representation, then $m_\p$ is a \textit{faithful abstract representation} of physical system $\p$ for the evolutions $C_\mathcal{T}(m_\p)$ and $\mathbf{H(p)}$.

The existence of such $\epsilon$-commuting diagrams define what is meant by a faithful abstract representation of a physical system. The final state of a physical object undergoing time evolution can be known \textit{either} by tracking the physical evolution and then representing the output abstractly, \textit{or} by evolving the abstract representation of the system. In the first case, the lower path of the diagram is followed; in the latter, the upper path. Finding out which diagrams $\varepsilon$-commute is the business of basic experimental science; once commuting diagrams have been established they can be exploited through engineering and technology. 

\subsection{Compute cycle}

\begin{figure}[t]
\centering
    \scalebox{0.7}{\begin{tikzpicture}[font=\large]

\draw[style=loosely dashed] (0,0.3) -- (11,0.3);

\node[physfront] (p) at (3,-1) {$\strut\p$};
\node[absfront] (mp) at (3,2) {$\strut m_{\p}$};
\draw[->,double] (mp) -- node[right, near start] {$\IRT$} (p);

\node[absfront] (mprp) at (8,2) {$\strut m^\prime_\p =_\varepsilon m_{\p^\prime}$};
\draw[absabsfront] (mp) -- node[above] {$C_\theory(m_\mathbf{p})$} (mprp);

\node[absback] (c) at (1,3.5) {$\strut A$};
	\draw[->] (c) to[out=0,in=90] (mp);
	\draw (3.5,3) node {\normalsize encode};

\node[absback] (cpr) at (10,3.5) {$\strut A^{\prime}$};
	\draw[->] (mprp) to[out=90,in=180] (cpr);
	\draw (7.5,3) node {\normalsize decode};

\node[physfront] (ppr) at (8,-1) {$\strut\p^\prime$};
\draw[physphysfront] (p) -- 
        node[above] {\normalsize program runs}
        node[below] {$\mathbf{H}(\p)$}
        (ppr);

\draw[->,double] (ppr) -- node[right, near start] {$\RT$} (mprp);
	
\end{tikzpicture}}
\caption{Physical computing in ART. 
An abstract problem $A$ is encoded into the model $m_{\p}$;
the model is instantiated into the physical computer state $\p$; 
the computer calculates via $\mathbf{H}(\p)$, evolving into physical state $\p^{\prime}$;
the final state is represented as the final abstract model $m_{\p^{\prime}} =_\varepsilon m^{\prime}_{\p}$;
this is decoded as the solution to the problem, $A^{\prime}$.
The instantiation, physical evolution, and representation together implement the desired abstract computation $C(m_\p)$.
(From now on we omit the dashed line separating the physical and abstract world, and rely on the different shaped boxes to indicate what components lie in which domain.)
}\label{computecy}
\end{figure}
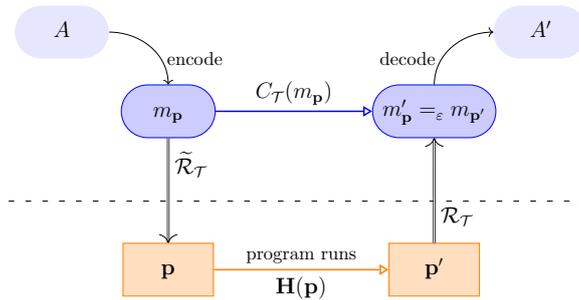

Figure~\ref{physabst} shows the basic \textit{science cycle}, of representing a physical system, and determining whether $C_\mathcal{T}$ is a sufficiently good abstract description of its behaviour, by requiring that $|m_\p - m_{\p^\prime}| \le \varepsilon$ for a sufficient range of initial states $\p$ to have confidence in $C_\mathcal{T}$.
There are derived variants of this diagram that capture the \textit{engineering cycle}, and the related \textit{compute cycle}.  See the original references for details; here we focus on the compute cycle (figure~\ref{computecy}).

An $\epsilon$-commuting diagram in the context of \textit{computation} also connects the physical computing device, $\p$, and its abstract representation $m_\p$. But to do so it makes use of the \textit{instantiation} relation $\mathcal{\widetilde{R}_\mathcal{T}} : M \rightarrow \bigp$.  Here, instead of saying abstract object $m_\p$ represents physical system $\p$, we say that physical system $\p$ instantiates abstract object $m_\p$.  Whereas the representation relation is primitive, the instantiation relation is a derived relation, based on multiple science cycles, abbreviated as $\mathcal{\widetilde{R}_\mathcal{T}}$; see original references for full details.

The use of $\mathcal{\widetilde{R}_\mathcal{T}}$ acknowledges that a computer is a physical system engineered to have a particular behaviour, rather than a natural physical system being scientifically modelled.  The full compute cycle is shown in figure~\ref{computecy}, starting from an initial abstract problem, through instantiation into a physical computer, physical evolution of the device, followed by representation of the final physical state as the abstract answer to the problem.

Ensuring that the diagram $\epsilon$-commutes is a process of debugging the physical system, including how it is instantiated (programmed and provided with input data), and how its output is represented.
This shows another key difference from the science cycle: there the diagram is made to $\epsilon$-commute by debugging the abstract model.

%

The most important use of a computing system is when the abstract outcome $m_\p$ is unknown: when computers are used to solve problems. Consider as an example the use of a computer to perform the binary arithmetical problem $01+10$. If the outcome were unknown, and the computing device being used to compute it, the final abstract state, $m^\prime_\p=(11)$, would not be evolved abstractly. Instead, confidence in the technological capabilities of the computer would enable the user to reach the final, abstract, output state $m_{\p^\prime} =_{\epsilon =0} m^\prime_\p$ using the physical evolution of the computing device alone. 
This use of a physical computer is the compute cycle, figure~\ref{computecy}: \textit{the use of a physical system} (the computer) \textit{to predict the outcome of an abstract evolution} (the computation).

Nothing in the above definition says that the physical computer has to be digital, or electronic, or universal, or pre-existing.  It could be a continuous analogue device; it could be a mechanical or organic device; it could be a hard-wired device with limited capabilities; it could be a one-shot device constructed for a particular computation.  It simply needs to be sufficiently powerful,  sufficiently accurate, and instantiable, to perform the RE's desired computations: the relevant squares must exist, and must $\varepsilon$-commute for the desired computations.
To demonstrate computation according to AR/T
it is sufficient to identify a particular set of AR/T components that establish $\epsilon$-commuting diagrams.

\subsection{The representational entity}

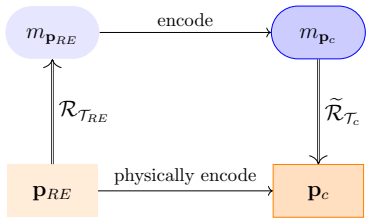
\begin{figure}[tp]
	\centering
 \scalebox{0.7}{\begin{tikzpicture}[font=\large]

\node[physback] (p) at (3,-1) {$\strut\p_{RE}$};
\node[absback] (mp) at (3,2) {$\strut m_{\p_{RE}}$};
\draw[->,double] (p) -- node[right] {$\RTRE$} (mp);
    
\node[absfront] (mprp) at (8,2) {$\strut m_{\p_c}$};
\draw[absabsside] (mp) -- node[above] {\normalsize encode} (mprp);

\node[physfront] (ppr) at (8,-1) {$\strut\p_c$};
\draw[physphysside] (p) -- node[above] {\normalsize physically encode} (ppr);
\draw[->,double] (mprp) -- node[right] {$\IRTc$} (ppr);

\end{tikzpicture}}
\caption{\label{fig:repn-a}The relationship between 
the physical representational entity $\p_{RE}$
and the physical computer $\p_{C}$
via abstract models of each.
There is an {\it encoding} of the abstract model $m_{\p_{RE}}$ into $m_{\p_{C}}$.
In a correctly working system,
this encoding is appropriately implemented by the respective physical systems: the square should $\varepsilon$-commute. 
Note that the models of the RE and the computer are potentially with respect to different theories.
} 
\end{figure}

As mentioned above, the representational entity (RE) supports the representation relation $\mathcal{R}$.
Although it does not appear explicitly in figure~\ref{computecy}, it is the physical entity that `owns' the abstract problem $A$ and desires the abstract solution $A'$. 
One issue glossed over in our original descriptions of ART that becomes crucial when analysing control systems is the relationship between the RE and the compute cycle.  
Here we summarise how the physical RE and its abstract state relate to the compute cycle  (figure~\ref{fig:repn-a}); see \cite{StepneyKendon2019} for full details. 

The RE's physical state is $\p_{RE}$.
The relevant part of the RE's physical state is represented as the abstract model $A = m_{\p_{RE}}$.  This is (our model of) the RE's problem. 
This model may incorrectly capture the RE's physical state, in which case the model needs to be modified; $m_{\p_{RE}}$ is our \textit{scientific} model of ${\p_{RE}}$. 

There is also the abstract model $m_{\p_{c}}$ that forms 
the specification of the encoding of the RE's problem as a computational problem. (This is the model $m_\p$ in figure~\ref{computecy}.)
The computers physical state may incorrectly implement this model, in which case, the physical state needs to be modified; $m_{\p_{c}}$ is an engineering model of ${\p_{c}}$.  
The RE's problem $m_{\p_{RE}}$ is \textit{encoded} into the computational model $m_{\p_{c}}$. 
There is no guarantee that such an encoding is possible: not all problems are computable.

The two representation/instantiation relation arrows in figure~\ref{fig:repn-a} are with respect to two different theories.
The representation ${\cal R}_{T_{RE}} : {\p_{RE}} \rightarrow m_{\p_{RE}}$ is based on the theory of how the physical RE forms abstract problem specifications;
the instantiation $\widetilde{\cal R}_{T_{c}} : m_{\p_{c}} \rightarrow {\p_{c}}$ is based on the theory of how the physical computer implements abstract computations.

In a correctly implemented computer, the diagram in figure~\ref{fig:repn-a} should $\varepsilon$-commute:
the instantiated state of the physical computer should correctly mirror the desired state of the physical RE: it should \textit{physically encode} the desired state.
The establishment of this physical encoding link is part of the engineering process.

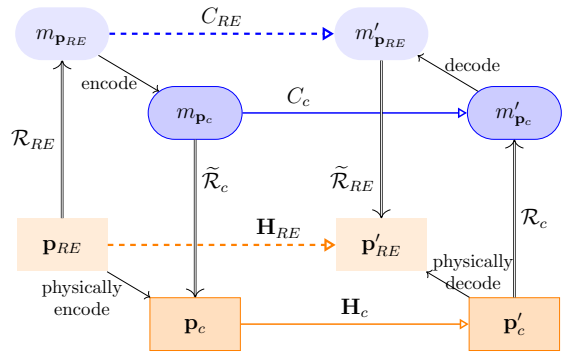
\begin{figure}[tp]
	\centering
 \scalebox{0.7}{\begin{tikzpicture}[font=\large]


\node[physfront] (p) at (3,-1) {$\strut\p_{c}$};
\node[absfront] (mp) at (3,3) {$\strut m_{\p_{c}}$};
\draw[->,double] (mp) -- node[right, near start] {$\IRc$} (p);
    
\node[absfront] (mprp) at (9,3) {$\strut m'_{\p_c}$};
\draw[absabsfront] (mp) -- node[above, near start] {$C_c$} (mprp);

\node[physfront] (ppr) at (9,-1) {$\strut\p'_c$};
\draw[physphysfront] (p) -- node[above] {${\mathbf H}_c$} (ppr);

\draw[->,double] (ppr) -- node[right] {$\Rc$} (mprp);


\node[physback] (p2) at (0.5,0.5) {$\strut\p_{RE}$};
\node[absback] (mp2) at (0.5,4.5) {$\strut m_{\p_{RE}}$};
\draw[->,double] (p2) -- node[left] {$\RRE$} (mp2);
    
\node[absback] (mprp2) at (6.5,4.5) {$\strut m'_{\p_{RE}}$};
\draw[absabsback] (mp2) -- node[above] {$C_{RE}$} (mprp2);

\node[physback] (ppr2) at (6.5,0.5) {$\strut\p'_{RE}$};
\draw[physphysback] (p2) -- node[above,near end]{${\mathbf H}_{RE}$} (ppr2);
\draw[->,double] (mprp2) -- node[left, near end] {$\IRRE$} (ppr2);


\draw[physphysside] (p2) -- node[left, at end,xshift=-4mm]{physically\\encode} (p);
\draw[absabsside] (mp2) -- node[left, near end]{encode} (mp);
\draw[physphysside] (ppr) -- node[right, at end]{physically\\decode} (ppr2);
\draw[absabsside] (mprp) -- node[right, near end,xshift=1mm] {decode} (mprp2);
	
\end{tikzpicture}}
\caption{\label{fig:repn-b}The full compute cycle including the representational entity and the physical computer.
The desired change in the RE's state, from posed problem to perceived solution, is ${\p_{RE}}\rightarrow {\p_{RE}}$.
The physical computer performs ${\p_{C}}\rightarrow {\p_{C}}$.
The full compute cycle from ART is:
we model the RE's physical state $\p_{RE}$ (desired computation) as abstract model $m_{\p_{RE}}$;
this is \textit{encoded} into the computational model $m_{\p_c}$;
the computation model is \textit{instantiated} into the physical computer state $\p$;
the physical computer \textit{evolves} to its final state $\p$;
the final physical state is \textit{represented} as the abstract computational solution $m_{\p_c}$;
the solution is \textit{decoded} to our model of the RE's final abstract problem solution $m_{\p_{RE}}$;
which is also (given $\epsilon$-commutation) our model of the instantiation of the final state of the RE.
Each set of squares (between representational entity and physical computer, and across the compute cycle) should  $\epsilon$-commute.
} 
\end{figure}

We now add this physical RE layer to the full compute cycle, 
figure~\ref{fig:repn-b}.
The full compute cycle involves traversal of many faces and edges of the displayed cube.
Each face has its own place in the model.
The back face is the RE's view of the computation;
the front face is the original compute cycle.
The left face is the encoding of the problem;
the right face is the decoding of the solution.
The top face is (our model of) the abstract performance of a computation by an abstract RE;
the bottom face is the physical use of a computer by a physical RE.

All of these relationships must be correctly implemented and modelled (the relevant squares containing encoding, decoding, instantiation, and representation must $\varepsilon$-commute) for the actual physical RE final state ${\p_{RE}}$ to be the desired physical RE final state, that is, for the physical computer to have been used correctly, and for it to have  performed correctly, to solve the RE's problem.
See \cite{StepneyKendon2019} for application of this model to the case of intrinsic computing in bacteria.

Nothing in the above definition says that the
RE has to be a brain, or even organic.
Here we demonstrate how an RE can appear in a control system loop as an engineered `proxy' for the ultimate RE: the controlled plant as proxy RE for the plant owner.

\section{Control systems}\label{sec:ctrlsys}
\subsection{Classical definition and models}

Here we provide a high level overview of the structure of classical control systems.
For more detail, see any standard textbook on the subject,
for example, \cite{Nise2024}.

A closed loop control system regulates the behaviour of a target system (the \textit{Plant}) via a control loop.  A continuous feedback control system compares the value or status of the process variable (output \textbf{\textsf y}) being controlled with the desired reference value or set-point (input reference \textbf{\textsf r}), and uses the difference to bring the process variable output to the same value as \textbf{\textsf r}.

\tikzset{
  serial-bkgd/.pic ={
      \fill[fill = orange!25, 
        rounded corners=0.6cm
      ] (-2,-2) rectangle ++(11.5,3) ;
      \fill[fill = white, 
        rounded corners=0.6cm
      ] (5.5,-1) rectangle ++(3,2.5) ;
      \fill[fill = white, 
      ] (8,-0.1) rectangle ++(4,1.15) ;
  }
}
\tikzset{
  pics/serial/.style 2 args ={
    code={
      \node (sum) [circle, 
            draw,
            minimum size=2mm,
            fill = white,
            ]
        {};
      \node (plus) at ($(sum) + (-0.2,0.35)$) {\textbf{+}} ; 
      \node (minus) at ($(sum) + (0.3,-0.3)$) {\textbf{--}} ;
    
      \node (control) [ rectangle,
        draw,
        fill = white,
        minimum width = 2.2cm,
        minimum height = 1cm,
        align = center      
      ] at ($(sum) + (3,0)$) {#1} ;
    
    \node (plant) [ rectangle,
        draw,
        fill = white,
        minimum width = 2.2cm,
        minimum height = 1cm,
        align = center
      ] at ($(control) + (4,0)$) {#2} ;

    \coordinate (start) at ($(sum) + (-1.8,0)$) ;
    \coordinate (feedback) at ($(plant) + (2,0)$) ;
      \draw[->, thick] (feedback) 
            -- ++(0,-1.5) 
            -- ++(-8,0)
            -| (sum) ;
      
      \draw[->, thick] (start) -- (sum) ;
      \draw[->, thick] (sum) -- (control) ;
      \draw[->, thick] (control) -- (plant) ;
      \draw[->, thick] (plant) -- ++(3,0) ;
  }
  }
}

\tikzset{
  parallel-bkgd/.pic ={
      \fill[fill = orange!25, 
        rounded corners=0.6cm
      ] (-2,-2.5) rectangle ++(7.5,3.5) ;
      \fill[fill = white, 
        rounded corners=0.6cm
      ] (1.5,-0.8) rectangle ++(3,2.5) ;
      \fill[fill = white, 
      ] (4,-0.1) rectangle ++(2,1.15) ;
  }
}
\tikzset{
  pics/parallel/.style 2 args ={
    code={
      \node (sum) [circle, 
            draw,
            minimum size=2mm,
            fill = white,
            ]
        {};
    \node (plus) at ($(sum) + (-0.2,0.35)$) {\textbf{+}} ; 
    \node (minus) at ($(sum) + (0.3,-0.3)$) {\textbf{--}} ;
    
    \node (plant) [ rectangle,
        draw,
        fill = white,
        minimum width = 2.2cm,
        minimum height = 1cm,
        align = center      
      ] at ($(sum) + (3,0)$) {#2} ;

    \node (control) [ rectangle,
        draw,
        fill = white,
        minimum width = 2.2cm,
        minimum height = 1cm,
        align = center      
      ] at ($(plant) + (0,-1.5)$) {#1} ;

    \coordinate (start) at ($(sum) + (-1.8,0)$) ;
    \coordinate (feedback) at ($(plant) + (2,0)$) ;
      \draw[->, thick] (feedback) |- (control) ;
      \draw[->, thick] (start) -- (sum) ;
      \draw[->, thick] (sum) -- (plant) ;
      \draw[->, thick] (plant) -- ++(3,0) ;
      \draw[->, thick] (control) -| (sum) ;
  }
  }
}

There are two standard formulations of feedback control systems, \textit{Serial} and \textit{Parallel}, see figures~\ref{fig:typeA} and~\ref{fig:typeB}.  
These are equivalent in the sense that one can transform each into the other; transforming from the Parallel to the Serial form adds a pre-filter before the summing junction.

\begin{figure}[tp]
\centering

\scalebox{0.7}{\begin{tikzpicture}
\tikzstyle{every node}=[font=\large\sffamily]
\pic {serial={Controller}{Plant}} ;

\node (r) at ($(start) + (0.3,0.3)$) {\textbf{r}} ;
\node (e) at ($(sum) + (1,0.3)$) {\textbf{e}} ;
\node (s) at ($(control) + (2,0.3)$) {\textbf{s}} ;
\node (y) at ($(plant) + (2.5,0.3)$) {\textbf{y}} ;

\end{tikzpicture}}

\caption{\label{fig:typeA}Control system diagram: Serial (or Cascade) Controller, operating on error signal \textbf{\textsf{e}},
where:
\textbf{\textsf{r}} is the (desired, preset) reference input;
\textbf{\textsf{y}} is the actual observed output of the plant; 
the summing junction combines these as \textbf{\textsf{r}$-$\textsf{y}} to give the error \textbf{\textsf{e}} input to the Controller;
\textbf{\textsf{s}} is the control signal input to the Plant, designed to reduce the error.
}
\centering

\scalebox{0.7}{\begin{tikzpicture}
\tikzstyle{every node}=[font=\large\sffamily]
\pic {parallel={Controller}{Plant}} ;

\node (r) at ($(start) + (0.3,0.3)$) {\textbf{r}} ;
\node (s) at ($(sum) + (1,0.3)$) {\textbf{s}} ;
\node (c) at ($(control) + (-2,0.3)$) {\textbf{c}} ;
\node (y) at ($(plant) + (2.5,0.3)$) {\textbf{y}} ;

\end{tikzpicture}}

\caption{\label{fig:typeB}Control system diagram: Parallel Controller, operating on output signal \textbf{\textsf{y}}, where:
\textbf{\textsf{y}} is the actual observed output of the plant, input to the Controller;   
\textbf{\textsf{c}} is the control signal; 
\textbf{\textsf{r}} is the (desired, preset) reference input;
the summing junction combines these as \textbf{\textsf{r}$-$\textsf{c}} to give the signal \textbf{\textsf{s}} input to the Plant, designed to reduce the error.
}
\end{figure}

In an open loop system, the human operator acts as the controller by adjusting the control signal to the Plant by hand.  
For example, in a basic aircraft, the pilot is the open-loop controller, changing the positions of the control surfaces, either through mechanically-mediated transduction, or through computer-mediated transduction.  If the aircraft is flying on autopilot, holding a height or speed, for example, it becomes a closed loop control system.

A classic example of an open loop control system is a toaster.
The user adjusts a timer \textit{by experience} to produce toast the way they like it.
It is an open loop system: there is no relationship between the input (length of time) and the output (darkness of toast) as the toast is cooking. There is no feedback, no comparison of toast colour against a desired colour, no computation of any kind, during a single operation of the device.
Fresh and stale bread will give different results. This motivates closed loop control.
Feedback can be used to account for variability of materials to provide a consistent product.
It could be achieved by an open loop system, where the user actively monitors the toast, and stops the device at the desired point. A closed loop system achieves this by monitoring toast colour automatically.

In a `bang-bang' control system there is no gradation of control signal: it is either on or off. 
Control is effected through setting the length of time the switch is on or off. In a central heating system, for example, the heating pump is either on or off.

From the outside, open loop, closed loop mechanical, and closed loop digital control systems are indistinguishable.
So the distinction of human versus mechanical versus digital is not sufficient to determine whether a control system is computing or not.

\subsection{Translating to ART}\label{sec:translate}

A note on notation.
We use three different typefaces to help distinguish the three different kinds of systems being described:
(i) the control system models as in figure~\ref{fig:typeABcombo}: sans serif font such as $\mathsf{y}, \mathsf{s}$;
(ii) the physical ART layer as in figure~\ref{fig:unwound} later: serif bold, such as $\mathbf{y}, \mathbf{s}$;
(ii) the abstract ART layer as in figure~\ref{fig:unwound-abs} later: serif italic, such as $y, s$.

\subsubsection{The proposed ART components}

The first step in translating the classical model of control systems into ART  concepts, in order to identify and locate any computing occurring, is to propose which parts of the system correspond to which ART components;
see figure~\ref{fig:typeABcombo}.
We then go on to show that these proposed components do indeed form
ART $\epsilon$-commuting diagrams for computation,
and hence the proposed identifications are correct,
and the controller is computing.

\begin{figure}[tp]
\centering

(a) \scalebox{0.5}{\begin{tikzpicture}
\tikzstyle{every node}=[font=\large\sffamily]
\pic {serial-bkgd};
\pic {serial={Controller}{Plant}} ;

\node (r) at ($(start) + (0.3,0.3)$) {\textbf{r}} ;
\node (e) at ($(sum) + (1,0.3)$) {\textbf{e}} ;
\node (s) at ($(control) + (2,0.3)$) {\textbf{s}} ;
\node (y) at ($(plant) + (2.5,0.3)$) {\textbf{y}} ;

\end{tikzpicture}}
(b) \scalebox{0.5}{\begin{tikzpicture}
\tikzstyle{every node}=[font=\large\sffamily]
\pic {parallel-bkgd};
\pic {parallel={Controller}{Plant}} ;

\node (r) at ($(start) + (0.3,0.3)$) {\textbf{r}} ;
\node (s) at ($(sum) + (1,0.3)$) {\textbf{s}} ;
\node (c) at ($(control) + (-2,0.3)$) {\textbf{c}} ;
\node (y) at ($(plant) + (2.5,0.3)$) {\textbf{y}} ;

\end{tikzpicture}}

\caption{\label{fig:typeABcombo}Control system diagrams: (a) Serial  and (b) Parallel Controllers, identifying the two components that appear in the ART version.  The plant corresponds to the physical representational entity $\p_{RE}$; the shaded area comprising preset reference, summing junction, controller, and observation of output, corresponds to the physical computer $\p_c$.
}
(a)~\scalebox{0.7}{\begin{tikzpicture}[font=\sffamily\large]

\node (pp) [ rectangle,
        draw,
        fill = white,
        minimum width = 2.2cm,
        minimum height = 1cm,
        align = center      
      ] at (0,0) {Plant} ;

\node (pc) [ rectangle,
        draw,
        fill = white,
        minimum width = 2.2cm,
        minimum height = 1cm,
        align = center      
      ] at (0,-1.5) {Controller} ;

    \coordinate (start) at ($(pp) + (-2.5,0)$) ;
    \coordinate (feedback) at ($(pp) + (2,0)$) ;

    \draw[->, thick] (pp) -- ++(3,0) ;
    \draw[->, thick] (feedback) |- (pc) ;
    \draw[->, thick] (pc) -| (start) -- (pp) ;

    \node (s) at ($(start) + (0.75,0.3)$) {\textbf{s}} ;
    \node (y) at ($(pp) + (2.5,0.3)$) {\textbf{y}} ;
\end{tikzpicture}}
\hspace{2cm}
(b)~\scalebox{0.7}{\begin{tikzpicture}[font=\large]

\node[physfront] (pc0) at (3,-1) {$\strut\p_{c}(t)$};
\node[physback] (pp0) at (3,1.5) {$\strut\p_{p}(t)$};
    
\draw[->] ([xshift=-2mm]pc0.north) 
    -- node[left] {$\textbf{s}(t)$}
    ([xshift=-2mm]pp0.south);
\draw[->] ([xshift=2mm]pp0.south) 
    -- node[right] {$\textbf{y}(t)$}
    ([xshift=2mm]pc0.north);
	
\end{tikzpicture}}
\caption{\label{fig:typeAB-ART}Proposed ART components of the Serial and Parallel Controllers.  
(a) Combining the control system components as indicated in figure~\ref{fig:typeABcombo}.
The Plant is unchanged; the summing junction and the reference input value \textbf{r} are folded into the box labelled Controller.
(b) Observing this system at a single timepoint $t$, in ART terms.
The physical plant $\p_{p}$ has physical output $\mathbf{y}$; 
this is observed by the physical controller $\p_{c}$;
the physical controller sends a physical signal $\mathbf{s}$ to the plant.
}
\end{figure}

We identify the  physical plant as the `user' of the control system.
The plant uses the controller to keep itself regulated;
it uses the controller to instruct it, through output $\mathbf{y}$ and signal $\mathbf{s}$, what to do in the next time period in order to maintain its desired behaviour.
So we propose to identify the physical plant as the \textit{physical representational entity}, which we here call $\p_p$ (subscript $p$ for `plant').

We identify the other components -- the reference input value \textbf{\textsf{r}}, summing junction and the controller -- as the \textit{physical computer}, $\p_c$.
We assume that the  reference input value is encoded in the state of the physical computer $\p_c$; we assume that changing the value of \textbf{\textsf{r}} is a separate operation.
We assume that any output or signal transduction is done within the physical controller;
this allows us also to capture any computation done as a part of signal transduction.


We next redraw the control system style diagram figure~\ref{fig:typeABcombo} as a diagram using ART concepts, figure~\ref{fig:typeAB-ART}b.
The two different control system figures have collapsed to a single ART concept figure,
via the intermediate control system figure~\ref{fig:typeAB-ART}a.
This shows the communications between the physical plant (representational entity) and the physical controller (computer).
Any signal transduction needed to convert the physical output $\mathbf{y}$ into the representation used in $\p_{c}$, or to produce the signal $\mathbf{s}$ into a form suitable for $\p_{p}$, is incorporated in $\p_{c}$.

\subsubsection{Unwinding the feedback}\label{sec:unwinding}

\begin{figure}[tp]
\centering
\scalebox{0.7}{\def\pplast{8}
\begin{tikzpicture}[font=\large]
\node (pp0) at (0.5,1.5) {} ;  
\node (pc0) at (0.5,0) {} ;
\foreach \i [remember=\i as \j (initially 0)] in {1,2,...,\pplast} {
    \node[rectangle, fill=orange!15, minimum size=0.5cm] (pp\i) at (\i*1.5,1.5) {};
    \node[rectangle, color=orange,draw, fill=orange!25, minimum size=0.5cm] (pc\i) at (\i*1.5,0) {};

    \draw[->] ([xshift=-1mm]pc\i.north) -- ([xshift=-1mm]pp\i.south);
    \draw[->] ([xshift=1mm]pp\i.south) --  ([xshift=1mm]pc\i.north);
    \draw[physphysback] (pp\j) -- (pp\i) ;
    \draw[physphysfront] (pc\j) -- (pc\i) ;
}
\draw[physphysback] (pp\pplast) -- ($(pp\pplast)+(1,0)$) ;
\draw[physphysfront] (pc\pplast) -- ($(pc\pplast)+(1,0)$) ;

\node[rectangle,draw, dash dot, color=gray,
    anchor = south west, minimum size = 2.7cm
    ] (zoom) at ($(pc4)-(0.6,0.6)$) {} ;
\draw[->, double] (zoom) -- ($(zoom)+(0,-2)$) ;

\end{tikzpicture}}\\
\vspace{8mm}
\scalebox{0.7}{\begin{tikzpicture}[font=\large]

\node[physfront] (pc0) at (3,-1) {$\strut\p_{c}(t_0)$};
\node[physback] (pp0) at (3,2) {$\strut\p_{p}(t_0)$};
    
\draw[dash dot,->] ([xshift=-2mm]pc0.north) 
    -- node[left] {$\textbf{s}(t_0)$}
    ([xshift=-2mm]pp0.south);
\draw[->] ([xshift=2mm]pp0.south) 
    -- node[right] {$\textbf{y}(t_0)$}
    ([xshift=2mm]pc0.north);
    
\node[physback] (pp1) at (9,2) {$\strut\p_p(t_1)$};
\draw[physphysback] (pp0) -- node[above] {$\textbf{H}_p(\p_p(t),\textbf{s}(t_0))$} (pp1);

\node[physfront] (pc1) at (9,-1) {$\strut\p_c(t_1)$};
\draw[physphysfront] (pc0) -- node[above] {$\textbf{H}_c(\p_c(t),\textbf{y}(t_0))$} (pc1);
	
\draw[->] ([xshift=-2mm]pc1.north) 
    -- node[left] {$\textbf{s}(t_1)$}
    ([xshift=-2mm]pp1.south);
\draw[dash dot,->] ([xshift=2mm]pp1.south) 
    -- node[right] {$\textbf{y}(t_1)$}
    ([xshift=2mm]pc1.north);

\draw[physphysfront, dash dot] ([xshift=-10mm]pp0.west) -- (pp0.west);
\draw[physphysfront, dash dot] ([xshift=-10mm]pc0.west) -- (pc0.west);
	
\draw[physphysfront, dash dot] (pp1.east) -- ([xshift=10mm]pp1.east);
\draw[physphysfront, dash dot] (pc1.east) -- ([xshift=10mm]pc1.east);
	
\draw[->] ([xshift=8mm,yshift=-15mm]pc0.east) 
    -- node[above, near end] {time}
    ([xshift=-3mm,yshift=-15mm]pc1.west);
	
\end{tikzpicture}}
\caption{\label{fig:unwound}
Unwinding the feedback diagram from fig~\ref{fig:typeAB-ART}b over time.
The top figure shows the unwinding over several timepoints,
with communication between the plant and computer (black arrows) accompanied by time evolution of each (orange arrows).
The bottom figure zooms in on two consecutive timepoints.
The inner solid and dashed arrows show a single feedback cycle;
the outer dash-dotted arrows show the preceding and following cycles.
See text in section~\ref{sec:unwinding} for details.
}
\end{figure}
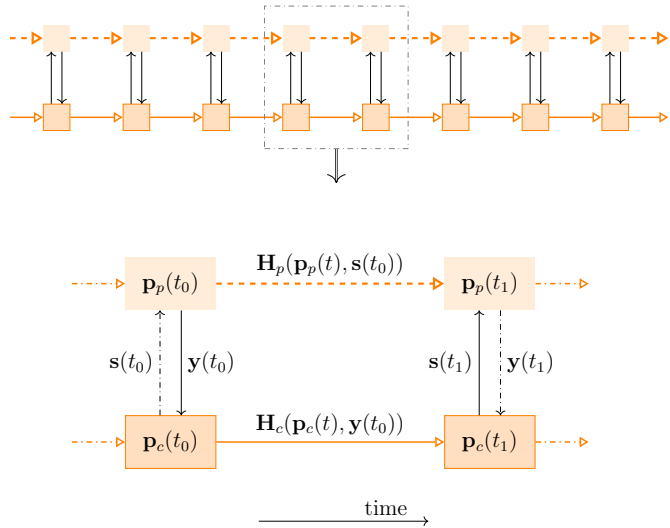

Although it uses ART concepts, figure~\ref{fig:typeAB-ART}b is a \textit{feedback} diagram, not an ART commuting square diagram.
The next step is to unwind the feedback, figure~\ref{fig:unwound}.
The top part of the figure shows the unwinding explicitly, giving the state of the system at consecutive timepoints (each being an instance of figure~\ref{fig:typeAB-ART}b),
connected by the time evolution of the plant and controller (orange arrows).
The bottom panel of figure~\ref{fig:unwound} zooms in on two consecutive timepoints, $t_0$ and $t_1$.
At time $t_0$, the physical plant $\p_p(t_0)$ sends its current output  $\mathbf{y}(t_0)$ to the physical controller $\p_c(t_0)$; the physical controllers state contains a representation of the previously programmed set point as $\mathbf{y}_{re}$.
The controller computes the next signal for the plant, using the plant output state at $t_0$; this takes the controller to state $\p_c(t_1) = \mathbf{H}_c(\p_c(t),\mathbf{y}(t_0))$; it sends the result of its computation, signal $\mathbf{s}(t_1)$, to the plant.
The plant, meanwhile, has been evolving under its own behaviour, along with the control signal it got from the preceding cycle, $\mathbf{s}(t_0)$.  When it receives the new signal, $\mathbf{s}(t_1)$, it is in state $\p_p(t_1) = \mathbf{H}_p(\p_p(t),\mathbf{s}(t_0))$.
So at the end of a single compute cycle (solid arrows), the physical plant has changed state to  $\p_p(t_1)$ due to its own evolution and the previous signal, has received a new signal $\mathbf{s}(t_1)$, the result of the computation, to use during its evolution for the next cycle,
and has sent its new output $\mathbf{y}(t_1)$ to the controller for its use in the next cycle.

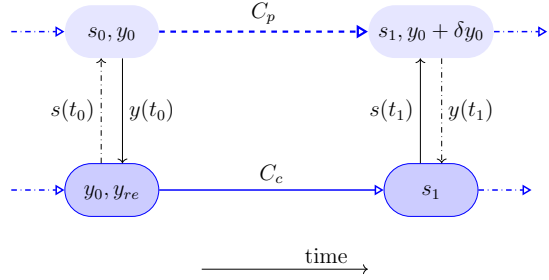
\begin{figure}[tp]
\centering
 \scalebox{0.7}{\begin{tikzpicture}[font=\large]

\node[absfront] (pc0) at (3,-1) {$\strut y_0, y_{re}$};
\node[absback] (pp0) at (3,2) {$\strut s_0, y_0$};
    
\draw[dash dot,->] ([xshift=-2mm]pc0.north) 
    -- node[left] {${s}(t_0)$}
    ([xshift=-2mm]pp0.south);
\draw[->] ([xshift=2mm]pp0.south) 
    -- node[right] {${y}(t_0)$}
    ([xshift=2mm]pc0.north);
    
\node[absback] (pp1) at (9,2) {$\strut s_1, y_0+\delta y_0$};
\draw[absabsback]  (pp0) -- node[above] {$C_p$} (pp1);

\node[absfront] (pc1) at (9,-1) {$\strut s_1$};
\draw[absabsfront] (pc0) -- node[above] {$C_c$} (pc1);
	
\draw[->] ([xshift=-2mm]pc1.north) 
    -- node[left] {${s}(t_1)$}
    ([xshift=-2mm]pp1.south);
\draw[dash dot,->] ([xshift=2mm]pp1.south) 
    -- node[right] {${y}(t_1)$}
    ([xshift=2mm]pc1.north);

\draw[dash dot,absabsfront] ([xshift=-10mm]pp0.west) -- (pp0.west);
\draw[dash dot,absabsfront] ([xshift=-10mm]pc0.west) -- (pc0.west);
	
\draw[dash dot,absabsfront] (pp1.east) -- ([xshift=10mm]pp1.east);
\draw[dash dot,absabsfront] (pc1.east) -- ([xshift=10mm]pc1.east);
	
\draw[->] ([xshift=8mm,yshift=-15mm]pc0.east) 
    -- node[above,near end] {time}
    ([xshift=-3mm,yshift=-15mm]pc1.west);
	
\end{tikzpicture}}
 \caption{\label{fig:unwound-abs}
The abstraction layer of the unwound feedback diagram figure~\ref{fig:unwound}.
At time $t_0$, the abstract plant sends its current abstract state $y_0$ as output  ${y}(t_0)$ to the abstract controller.
The controller computes the next signal for the plant, using the difference between observed and desired state $\Delta y_0 = y_0 - y_{re}$, where $y_0$ is the abstract plant output state at $t_0$ and $y_{re}$ is the previously programmed set point; it sends the result of its computation, $s_1(\Delta y_0)$, as signal ${s}(t_1)$, to the plant.  
When it receives this new signal, the abstract plant is in state $y_1 = y_0 + \delta y_0$; this change is due to it having been evolving under its own behaviour, under the influence of the control signal it received in the previous cycle, ${s}_0$.
 }
\end{figure}

This unwinding provides the ART-based physical model of the system, and it is this treatment of feedback  that is new to control systems when analysing physical computation in terms of ART.
In particular, the arrows indicating the plants evolution as RE during the feedback cycle (labelled $\mathbf{H}_p$ in the physical  figure~\ref{fig:unwound}, and labelled $C_p$ in the abstract  figure~\ref{fig:unwound-abs}) have a different interpretation
from those in the original compute cycle of figure~\ref{fig:repn-b}.
In the original cycle, these arrows show the desired computation, to be provided by the computer,
and there is no other RE evolution: the user simply waits, unchanged, for their result.
In the feedback cycle, however, the plant is also undergoing its own time evolution during this period,
and its desired computation is the \textit{correction}, or delta, to its evolved state.
Thus the computation being performed is not some simulation of the entire plant evolution,
but rather the relatively smaller, simpler computation of error correction.

Next, we abstract this unwinding of the physical layer into an abstract layer, and zoom in on a single cycle, figure~\ref{fig:unwound-abs}.
This is a minimal abstraction, capturing the minimal information used in simple control systems.
The abstract computation may need to exploit further information,
such as the past history of the state, in order to perform its calculations;
that extra information is omitted for clarity.

\subsection{Calculation, representation, and computing}\label{sec:ART-Control}

The final step of our proposed translation is to add 
the representation and instantiation relations to the compute cycle diagram.
If we are then able to form fully $\epsilon$-commuting compute-cycle diagrams with all components present, then we can conclude that the system is computing and our identifications are indeed correct.

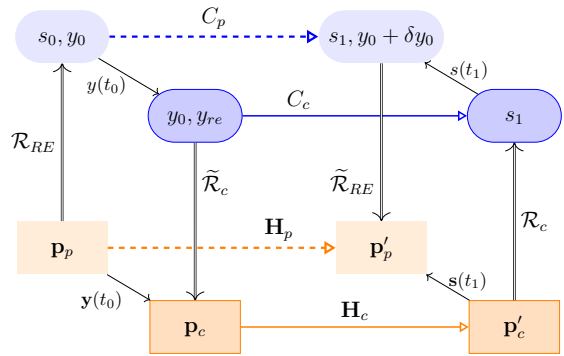
\begin{figure}[tp]
\centering
 \scalebox{0.7}{\begin{tikzpicture}[font=\large]


\node[physfront] (p) at (3,-1) {$\strut\p_{c}$};
\node[absfront] (mp) at (3,3) {$\strut y_0, y_{re}$};  
\draw[->,double] (mp) -- node[right,near start] {$\IRc$} (p);

\node[absfront] (mprp) at (9,3) {$\strut s_1$};
\draw[absabsfront] (mp) -- node[above, near start] {$C_c$} (mprp);

\node[physfront] (ppr) at (9,-1) {$\strut\p'_c$};
\draw[physphysfront] (p) --  node[above] {${\mathbf H}_c$} (ppr);
\draw[->,double] (ppr) -- node[right] {$\Rc$} (mprp);


\node[physback] (p2) at (0.5,0.5) {$\strut\p_{p}$};
\node[absback] (mp2) at (0.5,4.5) {$\strut s_0, y_0$};
\draw[->,double] (p2) -- node[left] {$\RRE$} (mp2);
    
\node[absback] (mprp2) at (6.5,4.5) {$\strut s_1, y_0 + \delta y_0$};
\draw[absabsback] (mp2) -- node[above] {$C_{p}$} (mprp2);

\node[physback] (ppr2) at (6.5,0.5) {$\strut\p'_{p}$};
\draw[physphysback] (p2) -- node[above, near end] {${\mathbf H}_{p}$} (ppr2);
\draw[->,double] (mprp2) -- node[left,near end] {$\IRRE$} (ppr2);


\draw[physphysside] (p2) -- node[left, at end,xshift=-4mm]{$\textbf{y}(t_0)$} (p);
\draw[absabsside] (mp2) -- node[left, near end,xshift=-2mm]{$y(t_0)$} (mp);
\draw[physphysside] (ppr) -- node[right, near end,xshift=1mm]{$\textbf{s}(t_1)$} (ppr2);
\draw[absabsside] (mprp) -- node[right, near end,xshift=1mm] {$s(t_1)$} (mprp2);
	
\end{tikzpicture}}
\caption{\label{fig:ART_ctrl}
Incorporating a single compute cycle of the unwound physical system from figure~\ref{fig:unwound} and unwound abstraction from figure~\ref{fig:unwound-abs} into the ART cycle of fig~\ref{fig:repn-b}.
This should be interpreted as a single cycle in the context of a sequence of cycles, including previous and subsequent signals and outputs, as shown explicitly in figures~\ref{fig:unwound} and~\ref{fig:unwound-abs}.
The controller $\textbf{p}_c$ plays the role of the \textit{physical computer}; the plant $\textbf{p}_p$ plays the role of the \textit{representational entity}: the entity on whose behalf the computation is being performed.
}
\end{figure}

Compare the physical layer shown in figure~\ref{fig:unwound}
and the abstract layer in figure~\ref{fig:unwound-abs}
with the generic representational entity figure~\ref{fig:repn-b}.
This leads us to the full control system in figure~\ref{fig:ART_ctrl},
including the abstract computation $C_c$ performed by the physical controller, 
and the abstract computation $C_p$ desired by the RE.
In general, we identify the preset, the summing junction, and the Controller
(contained in the shaded area in figure~\ref{fig:typeABcombo}; combined into $\textbf{p}_c$ in figure~\ref{fig:typeAB-ART}) as our device of interest.
We have \textit{calculation}:
even in the edge case of a system where there is no need for a specific Controller, there is always an implementation of the calculation
 $\textsf{r}-\textsf{y}$ or $\textsf{r}-\textsf{c}$ in the summing junction (figure~\ref{fig:typeABcombo}) as a minimum\footnote{%
This edge case is expected to be rare in control systems, however, as there is almost certainly some transduction occurring (in the Controller), since the type/dimension of \textsf{s}, the input to the Plant is unlikely to be the same as the type/dimension of \textsf{y}, the observed output from the Plant.
}.
Furthermore, we have \textit{representation} (physical input signal \textbf{s} and physical output \textbf{y} represented as abstract $s$ and $y$ respectively).  
There is always the need for representation, as the physical output \textbf{y} needs to be represented in order to be suitably transduced for input to the summing junction.

So, according to ART, the controller $\textbf{p}_c$ is computing.
Since this is a generic description of a control system, we claim that, according to ART, \textit{any control system is computing}.

\subsection{Feedback and error propagation}

There are two key steps in modelling control systems as computers using ART.

Firstly, we identify the plant with the representational entity,
the entity on whose behalf the computation is being performed.
Actually, the plant is a \textit{proxy} RE for the true RE:
the engineer who designed and calibrated the plant and its controller to work as desired.

Secondly, we unwind the feedback loop inherent in a control system
to get a sequence of $\epsilon$-commuting diagrams.
Here we see why the diagrams may only $\epsilon$-commute,
rather than perfectly compute:
the RE/Plant is autonomously changing its own state even as the physical computing is happening; it is no longer in the state it was in when it initiated the compute cycle.
The control calculation is not trivial; the result needs to avoid over-damping, under-damping, or instability of the output.
This is the core of controller design theory, and is not considered further here.
(For more detail, see, for example \cite{Nise2024}.)
For our purposes,
provided the timescales are short enough, and the calculation is well-designed,
then the errors are small enough that the supplied answer is sufficiently good for the purpose of controlling the Plant.
We do note that physical hysteresis may be employed 
to provide some damping
and as a form of memory in a non-digital control system, enabling it to determine if the plant output is increasing or decreasing, for example.

Traditional analogue computers can suffer from an accumulation of errors.
In a continuous control system, however, the feedback from the physical plant provides a reality check
that can be used to correct for the errors.

\section{Control systems examples in ART terms}\label{sec:control_ART}

Here we provide a series of examples of digital, mechanical, and human-in-the-loop, control systems; 
how they are modelled in ART;
and where the representation and computation occur.
These are all simple examples related to controlling power output in different ways.

\subsection{Digital control systems}\label{sec:dgital control}
In a digital control system, the controller is some form of digital computer.
\subsubsection{Example: a digital thermostat}\label{sec:digital eg}
\begin{figure}[tp]
\centering
{
\includegraphics[scale=0.1]{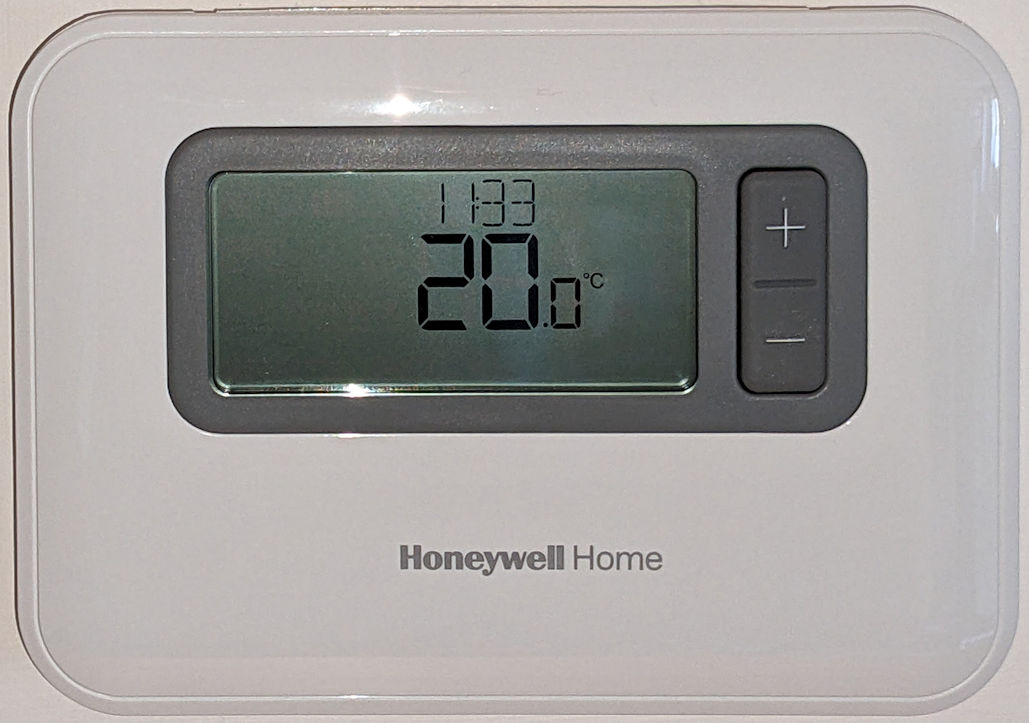}
}
\caption{\label{fig:digital_thermostat}
A simple digital thermostat.
[Photo credit: SS]
}
\end{figure}

Consider a simple digital thermostat,
that might be found on the wall in a house or hotel (see figure~\ref{fig:digital_thermostat}), that controls the room temperature by regulating the output of a heater (the plant).
These can typically be programmed to change the temperature set point at different times;
here we just consider their action at maintaining a fixed temperature.

The thermostat contains a temperature sensor,
which elicits the current room temperature.
If this is lower than the set temperature, the thermostat signals the heater to switch on. 
If the temperature is too high, the controller signals the heater to switch off.
Control theory is used to ensure sufficient damping, so that the heater does not constantly switch on and off when the room is at the set temperature.

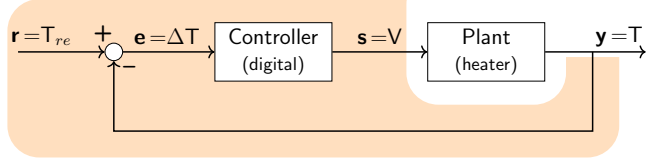
\begin{figure}[tp]
\centering
\scalebox{0.7}{\begin{tikzpicture}
\tikzstyle{every node}=[font=\large\sffamily]
\pic {serial-bkgd};
\pic {serial={Controller\\\normalsize(digital)}{Plant\\\normalsize(heater)}} ;

\node (r) at ($(start) + (0.3,0.3)$) {~~\textbf{r}\,=T$_{re}$} ;
\node (e) at ($(sum) + (1,0.3)$) {\textbf{e}\,=$\Delta$T} ;
\node (s) at ($(control) + (2,0.3)$) {\textbf{s}\,=V} ;
\node (y) at ($(plant) + (2.5,0.3)$) {\textbf{y}\,=T} ;

\end{tikzpicture}}
\caption{\label{fig:digital_heater}
The digital thermostat computer (shaded) as part of the overall Control system.
See section~\ref{sec:digital eg}  text for details.
}
\end{figure}

See figure~\ref{fig:digital_heater} 
for the control system diagram.
The reference baseline $\mathsf{r = T_{re}}$ is the preset desired temperature.
It has to be changed to modify the preset value, for example, by the keying in of a new temperature value. 
The difference between current observed $\mathsf{y=T}$ and set point $\mathsf{T_{re}}$ is computed as the error $\mathsf{e = \Delta T}$,
and the controller calculates the value of the voltage signal $\mathsf{s = V}$, to switch the heater on, to increase the temperature, or to switch off if the temperature is too high.

\subsubsection{The digital thermostat system in ART}\label{sec:digital-ART}

\begin{figure}[tp]
\centering
 \scalebox{0.7}{\begin{tikzpicture}[font=\large]


\node[physfront] (p) at (3,-1) {$\strut\p_{c}$};
\node[absfront] (mp) at (3,3) {$\strut T_0, T_{re}$};  
\draw[->,double] (mp) -- node[right,near start] {$\IRc$} (p);

\node[absfront] (mprp) at (9,3) {$\strut V_1$};
\draw[absabsfront] (mp) -- node[above, near start,xshift=2mm] {\normalsize compute signal} (mprp);

\node[physfront] (ppr) at (9,-1) {$\strut\p'_c$};
\draw[physphysfront] (p) --  node[above] {\normalsize digital controller} (ppr);
\draw[->,double] (ppr) -- node[right] {$\Rc$} (mprp);


\node[physback] (p2) at (0.5,0.5) {$\strut\p_{p}$};
\node[absback] (mp2) at (0.5,4.5) {$\strut V_0, T_0$};
\draw[->,double] (p2) -- node[left] {$\RRE$} (mp2);
    
\node[absback] (mprp2) at (6.5,4.5) {$\strut V_1, T_0 + \delta T_0$};
\draw[absabsback] (mp2) -- node[above] {\normalsize restore temp} (mprp2);

\node[physback] (ppr2) at (6.5,0.5) {$\strut\p'_{p}$};
\draw[physphysback] (p2) -- node[above, near end] {\normalsize heater} (ppr2);
\draw[->,double] (mprp2) -- node[left,near end] {$\IRRE$} (ppr2);


\draw[physphysside] (p2) -- node[left, at end,xshift=-4mm]{$\textbf{T}(t_0) =$\\room temp} (p);
\draw[absabsside] (mp2) -- node[left, near end,xshift=-2mm]{$T(t_0)$} (mp);
\draw[physphysside] (ppr) -- node[right, near end,xshift=0.5mm]{$\textbf{V}(t_1) =$ new voltage} (ppr2);
\draw[absabsside] (mprp) -- node[right, near end,xshift=1mm] {$V(t_1)$} (mprp2);
	
\end{tikzpicture}}
\caption{\label{fig:digital_ART}
The digital thermostat computer in  ART.
See section~\ref{sec:digital-ART} text for details.
}
\end{figure}
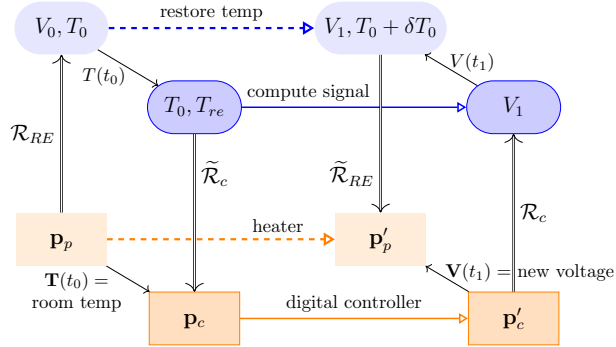

See figure~\ref{fig:digital_ART} 
for how this digital control system can be modelled in ART.

The components of the control system in figure~\ref{fig:digital_heater} become the physical base layer of the ART cube in figure~\ref{fig:digital_ART}.
The current room temperature $\mathbf{T}(t_0)$ is physically sensed and encoded as some bit pattern in the digital controller,
along with some encoding to instantiate the set value $\mathbf{T_{re}}$,
as part of its current physical state $\mathbf{p}_c$.
The controller calculates the difference between the encoded values of $\mathbf{T}(t_0)$ and $\mathbf{T_{re}}$, and the new voltage needed to reduce this difference, through the physical actions of its digital electronic components.
The calculated voltage bit pattern is physically decoded through some digital to analogue converter and transducer,
into the voltage $\mathbf{V}(t_1)$ applied to the heater in the next time step.

In the abstract top layer of the cube,
which is our model of the physical systems operation,
the plant's desired action is to restore the room temperature $T_0$ to the preset value $T_{re}$.
It outputs the current temperature $T_0$ for (abstract) computation, which uses the error $\Delta T_0 = T - T_{re}$, and receives back the computed signal of how to achieve this, $V_1(\Delta T_0)$.
Its final state is $T_1 = T_0 + \delta T_0$,
the temperature it managed to achieve using the previous signal;
this will not necessarily be the desired temperature $T_{re}$, but, in a well-designed control system, will be making progress towards it.

There is computation occurring in the digital controller evolution in figure~\ref{fig:digital_ART},
corresponding to the 
calculation of the control system values, error $\mathsf{e} = \Delta \mathsf{T}$ and signal $\mathsf{s}$, from figure~\ref{fig:digital_heater}.
There is representation occurring,
as the temperature $T, T_0$ and voltage $V$ variables
in the abstract part of the compute cycle are
abstract representations of physically instantiated bit patterns in the digital controller.

We have belaboured this description in the case of a digital controller,
where the presence of computation is well-understood,
in order to use the same structure for the less clear cases of the mechanical and open-loop control systems below.

\subsection{Electro-mechanical control systems}\label{sec:coil}
In an electro-mechanical control system, the controller is has both an electrical and a mechanical component realising its behaviour.
\subsubsection{Example: a binary electro-mechanical thermostat}\label{sec:coil eg}

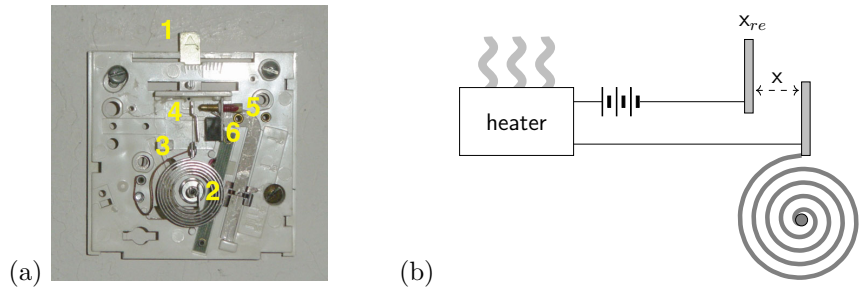
\begin{figure}[tp]
\centering
{(a)~\includegraphics[width=0.3\columnwidth]{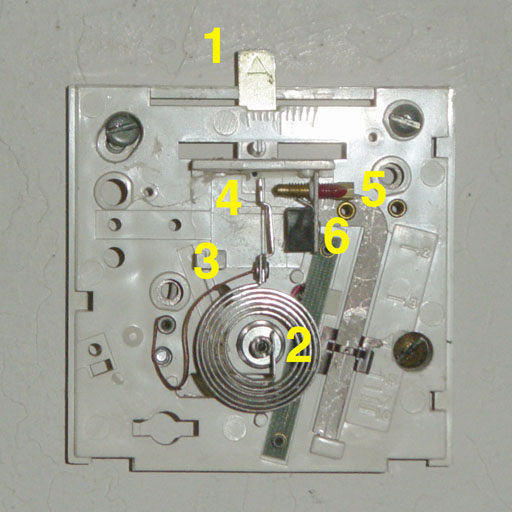}}
{
\qquad(b)~
\scalebox{0.75}{
\tikzset{fumes/.style={line width =1.5mm, color = gray!50, decorate, decoration={snake, segment length=6mm, amplitude=1mm}}}
\begin{tikzpicture}
\tikzstyle{every node}=[font=\large\sffamily]

\draw[line width =0.8mm, color = gray, 
    domain=90:360*5-90, samples=100,smooth,variable=\t]
  plot (-\t:\t/1500) ;

\draw[fill = gray] circle (3pt) ; 

\coordinate (c1) at ($(0, 1.14)$) ;
\draw[fill = gray!50] (c1) rectangle ++(0.15,1.3) ;

\coordinate (c2) at ($(c1) + (-1,0.75)$) ;
\draw[fill = gray!50] (c2) rectangle ++(0.15,1.3) ;
\node at ($(c2)+(0.15,1.3)+(0,0.3)$) {x$_{re}$};

\draw ($(c1) + (0,0.2)$) -- ($(c1) + (-6,0.2)$) ;
\draw ($(c2) + (0,0.2)$) -- ($(c2) + (-5,0.2)$) ;

\coordinate (heater) at ($(c1)+(-5,0.6)$) ;
\foreach \i in {-0.5,0,0.5} 
    \draw[fumes] ($(heater)+(\i,1.5)$) -- ($(heater)+(\i,0)$) ;
\node[
    draw,
    rectangle,
    minimum width = 2cm,
    minimum height = 1.2cm,
    fill = white,
] at (heater) {heater} ;

\coordinate (battery) at ($(c2)+(-2.5,0.2)$) ;
\fill[white] ($(battery) + (0,-0.1)$) rectangle ($(battery) + (0.65,0.1)$) ;
\foreach \i in {0,0.25,0.5} {
    \draw ($(battery) + (\i,-0.25)$) rectangle ++(0.01,0.5) ;
    \draw[fill = black] ($(battery) + (\i+0.1,-0.1)$) rectangle ++(0.05,0.2) ;
}

\draw[<->, dashed] ($(c2)+(0.2,0.4)$) -- node [above,midway] {x}  ($(c1)+(-0.05,1.15)$) ;

\end{tikzpicture}}
}
\caption{\label{fig:bimetallic}
(a) A bimetallic coil thermostat.
Lever (1) changes the set point.
Coil (2)  is clamped at the centre.
As the end moves, it can make contact with (4).
[image credit: Leonard G.  Public Domain, via Wikimedia Commons https://commons.wikimedia.org/wiki/File:WPThermostat.jpg] 
(b) Schematic of the bimetallic coil control system.  
$x_{re}$ is the set point (position of the contact);
$x$ is the distance from this (distance from the end of the coil),
and decreases as the temperature drops and the strip uncoils.
}
\end{figure}

Before there was digital electronics, there were still thermostats.
One design uses the different thermal expansion properties of different materials.
Two metals of different thermal expansion coefficients are bonded together to form a bimetallic strip.
When heated, one side expands more than the other, and the strip bends an amount dependent on the temperature.
This property is exploited to form a thermostat:
the bending is used to actuate a switch when the temperature drops to some set point,
to switch on a heater.
See figure~\ref{fig:bimetallic}.

\begin{figure}[tp]
\centering
{\scriptsize (a)}\scalebox{0.65}{\begin{tikzpicture}
\tikzstyle{every node}=[font=\large\sffamily]
\pic {parallel-bkgd};
\pic {parallel={Controller\\\normalsize(coil)}{Plant\\\normalsize(heater)}} ;

\node (r) at ($(start) + (0.3,0.3)$) {~~\textbf{r}\,=\,x$_{re}$} ;
\node (s) at ($(sum) + (1,0.3)$) {\textbf{s}\,=1/0} ;
\node (c) at ($(control) + (-2,0.3)$) {\textbf{c}\,=\,x} ;
\node (y) at ($(plant) + (2.5,0.3)$) {\textbf{y}\,=T} ;

\end{tikzpicture}}\\\vspace{3mm}
{~\scriptsize (b)}\scalebox{0.65}{\begin{tikzpicture}[font=\large]


\node[physfront] (p) at (3,-1) {$\strut\p_{c}$};
\node[absfront] (mp) at (3,3) {$\strut T_0, T_{re}$};  
\draw[->,double] (mp) -- node[right,near start] {$\IRc$} (p);

\node[absfront] (mprp) at (9,3) {$\strut s_1\in\{0,1\}$};
\draw[absabsfront] (mp) -- node[above, near start,xshift=2mm] {\normalsize compute signal} (mprp);

\node[physfront] (ppr) at (9,-1) {$\strut\p'_c$};
\draw[physphysfront] (p) --  node[above] {\normalsize bimetallic coil} (ppr);
\draw[->,double] (ppr) -- node[right] {$\Rc$} (mprp);


\node[physback] (p2) at (0.5,0.5) {$\strut\p_{p}$};
\node[absback] (mp2) at (0.5,4.5) {$\strut s_0, T_0$};
\draw[->,double] (p2) -- node[left] {$\RRE$} (mp2);
    
\node[absback] (mprp2) at (6.5,4.5) {$\strut s_1, T_0 + \delta T_0$};
\draw[absabsback] (mp2) -- node[above] {\normalsize restore temp} (mprp2);

\node[physback] (ppr2) at (6.5,0.5) {$\strut\p'_{p}$};
\draw[physphysback] (p2) -- node[above, near end] {\normalsize heater} (ppr2);
\draw[->,double] (mprp2) -- node[left,near end] {$\IRRE$} (ppr2);


\draw[physphysside] (p2) -- node[left, at end,xshift=-4mm]{$\textbf{T}(t_0) =$\\room temp} (p);
\draw[absabsside] (mp2) -- node[left, near end,xshift=-2mm]{$T(t_0)$} (mp);
\draw[physphysside] (ppr) -- node[right, near end,xshift=1.8mm]{$\textbf{s}(t_1) =$ on/off} (ppr2);
\draw[absabsside] (mprp) -- node[right, near end,xshift=1mm] {$s(t_1)$} (mprp2);
	
\end{tikzpicture}}
\caption{\label{fig:bimetallic_control}
(a) The bimetallic coil physical computer (shaded) as part of the overall Control system.
See section~\ref{sec:coil eg} text for details.
(b) The bimetallic coil computer in  ART.
See section~\ref{sec:coil ART} text for details.
}
\end{figure}
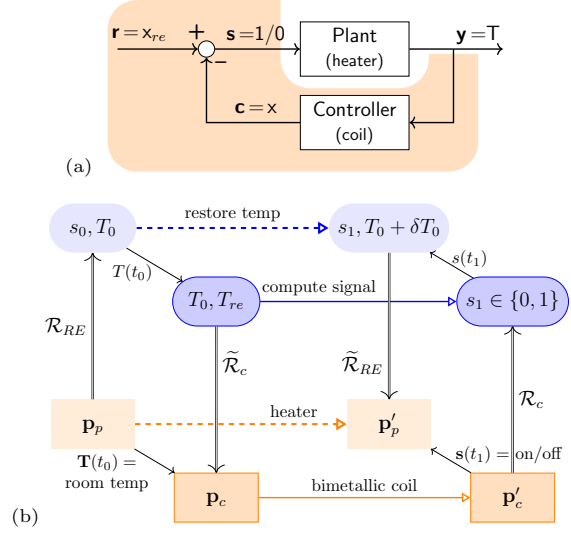

See figure~\ref{fig:bimetallic_control}a for the control system diagram.
The reference baseline $\mathsf{r = x_{re}}$ gives the preset desired temperature, calibrated so the contact position is the coil position at that temperature.  It has to be changed to modify the preset value (for example, by turning a dial or moving a lever to change the position of the movable contact).
The Controller is the coil, transducing physical temperature $\mathsf{T}$ to movement $\mathsf{x}$ through its physical unwinding.
The difference between positions $\mathsf{x_{re}}$ and $\mathsf{x}$ is computed in a \textit{boolean} summing junction: it is $\mathsf{0}$ if $\mathsf{x_{re}} < \mathsf{x}$ (not touching), and $\mathsf{1}$ if $\mathsf{x_{re}} = \mathsf{x}$ (touching).
This boolean signal $\mathsf{s = 0/1}$ switches the heater on or off.

\subsubsection{The electro-mechanical thermostat in ART}\label{sec:coil ART}

See figure~\ref{fig:bimetallic_control}b for how the bimetallic controller can be modelled in ART.

The components of the control system in figure~\ref{fig:bimetallic_control}a become the physical base layer of the ART cube in figure~\ref{fig:bimetallic_control}b.
The current room temperature $\mathbf{y}(t_0)$ is physically sensed by the bimetallic coil;
along with the encoding of the set value $x_0$,
this forms its current physical state $\mathbf{p}_c$.
The bimetallic coil reacts to the input temperature by increasing or decreasing its coiling,
moving its end relative to the set point contact,
forming its final physical state $\mathbf{p}_c$.
Whether this movement results in the switch state being changed directly signals whether the heater needs to be switched on or off.

The abstract top layer of the cube
is identical to the digital controller case, figure~\ref{fig:digital_ART},
except for the signal: here it is a boolean on/off, rather than a continuous voltage.
The very different physical implementations (digital controller versus bimetallic strip plus electric switch)
have been abstracted away, and the only difference visible abstractly is the type of the signal.

There is computation occurring in the bimetallic controller evolution in figure~\ref{fig:bimetallic_control}b,
corresponding to the 
calculation of the control system values, position $c = \mathsf{x}$ and boolean signal $\mathsf{s = 1/0}$, from figure~\ref{fig:bimetallic_control}a.
There is representation occurring,
as the temperatures $T_0, T_{re}$ and final state $s_1$ variables
in the abstract part of the compute cycle are
abstract representations of coil position $\mathbf{x}$ and contact making $\mathbf{s}$ in the physical controller.

In such a physical control system,
where the plant output $\mathsf{y}$
is directly sensed by the controllers physical properties (here, thermal expansion in the coil),
and the controllers physical properties directly signal the plant (here, the state to the controller switch controlling the heater on/off state),
there is no need for any explicit sensor input or output conversion,
unlike in the digital case, where the temperature sensor has to convert the sensed input into electrical digital signals, and the calculated voltage value has to be converted to a physical voltage.
Transduction happens automatically due to the designed physical properties of the controller.


\subsection{Mechanical control systems}\label{sec:guv}
Before there was electricity, there were mechanical control systems, 
where the controller is a purely mechanical device.
\subsubsection{Example: a continuous centrifugal governor}\label{sec:guv eg}

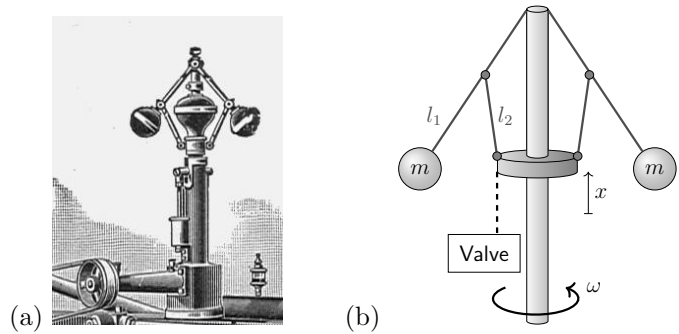
\begin{figure}[tp]
\centering
{(a)~\includegraphics[width=0.25\columnwidth]{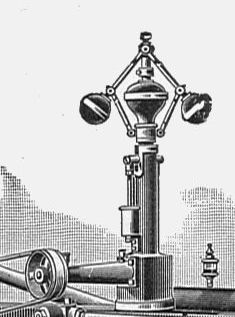}}
{\qquad(b)~
\scalebox{0.7}{\begin{tikzpicture}
\tikzset{every node/.style={line width = .15mm, font=\large\sffamily}
}
\begin{scope}[xshift = 7cm]

\node (shaft) [cylinder, 
    shape border rotate=90, 
    draw,
    minimum height=6cm,
    minimum width=4mm,
    fill = gray!20,
    aspect = 0.5,   
      left color=gray!5,
      middle color=gray!25,
      right color=gray!50,
      shading angle=90,
    ]
{};

\shade[ball color = gray!60, opacity = 0.4] ($(shaft.west) -(2,0)$) circle (0.4cm) ;
\node (mass1) at ($(shaft.west) -(2,0)$) [draw,
    shape = circle,
    minimum size = 0.8cm
] {$m$} ;
\shade[ball color = gray!60, opacity = 0.4] ($(shaft.east) +(2,0)$) circle (0.4cm) ;
\node (mass2) at ($(shaft.east) +(2,0)$) [draw,
    shape = circle,
    minimum size = 0.8cm
] {$m$} ;

\node (collar) [cylinder, 
    shape border rotate=90, 
    draw,
    minimum height=1mm,
    minimum width=1.5cm,
      left color=gray!25,
      middle color=gray!60,
      right color=gray!100,
      shading angle=90,
    ]
{};

\node (shaft1)  at ($(shaft.mid) + (0,0.12)$) [cylinder, 
    shape border rotate=90, 
    draw,
    minimum height=2.88cm,
    minimum width=4mm,
    fill = gray!20,
    anchor = bottom,
    aspect = 0.5,   
      left color=gray!5,
      middle color=gray!25,
      right color=gray!50,
      shading angle=90,
    ]
{};

\draw[|->] ($(collar.before bottom) + (0.2, -0.8)$) -- node[right]{$x$} ($(collar.before bottom) + (0.2, 0)$) ;

\draw[very thick, color=black!70] (shaft.before top) -- node[left, near end]{$l_1~$}  (mass1.55) ;
\draw[very thick, color=black!70] (shaft.after top) -- (mass2.125) ;

\coordinate (M1) at ($(shaft.before top)!0.45!(mass1.north east)$);
\draw[very thick, color=black!70] (M1) -- node[right]{$l_2$}  (collar.before top) ;
\coordinate (M2) at ($(shaft.after top)!0.45!(mass2.north west)$);
\draw[very thick, color=black!70] (M2) -- (collar.after top) ;

\draw[fill = black!50] (M1) circle (0.8mm) ;
\draw[fill = black!50] (M2) circle (0.8mm) ;
\draw[fill = black!50] (collar.before top) circle (0.8mm) ;
\draw[fill = black!50] (collar.after top) circle (0.8mm) ;

\node (valvepos) at ($(shaft.center) +(-1,-1.6)$) [draw] {} ;
\draw[dashed, very thick] (collar.after bottom) |- (valvepos) ;
\node (valve) at ($(shaft.center) +(-1,-1.6)$) [
    draw, 
    fill = white,
    inner sep=6pt
] {Valve} ;

\draw[->, very thick] ($(shaft.south) + (-0.6,0.7)$) arc (-225:45:0.8 and 0.3) node[right] {$~~\omega$} ;
\end{scope}
\end{tikzpicture}}
}
\caption{\label{fig:guv}
(a) The Centrifugal Governor.  [image credit: Andy Dingley (scanner). Scan from Nehemiah Hawkins (1904 edition of 1897 book). \textit{New Catechism of the Steam Engine}, Theo Audel.  Public Domain, via Wikimedia Commons https://commons.wikimedia.org/w/index.php?curid=9567390] 
(b) Schematic view. The whole apparatus rotates at the engine speed $\omega$.  As $\omega$ increases, the fly balls (mass $m$) raise; as it decreases, they drop under gravity.  As the fly balls raise and lower, the linkages $l_1$ and $l_2$ change the position of the grey collar, to a distance $x$ from some baseline; the height of the collar $x$ represents the rotation speed $\omega$.  The collar connects to a valve (dashed line: precise details of connection not shown here).
}
\end{figure}

Before there was electricity, there were purely mechanical control systems, for example, governors regulate the speed of a machine \cite{Maxwell-1868}.
The centrifugal governor (Figure~\ref{fig:guv}a) is a mechanical feedback control system that
was devised in the 17th century
to regulate the separation of flour millstones 
to avoid friction when high wind on the sails of the driving windmill risked turning the upper running stone too fast. 
This condition risked scorching of the grain or
causing an explosion of the very fine flour dust. 

It was later adapted by James Watt to regulate a steam engine 
under varying load conditions, by controlling the amount of applied steam. Under increased load, the engine would slow down, causing the flyballs to drop. Through a series of linkages, this movement caused the steam valve to open, thereby increasing torque applied to the load and consequently bringing the engine back up to the desired speed (figure~\ref{fig:guv}b).  
This uses the principle of proportional control. 
The original system uses gravity to bring the flyballs back close to the rotating shaft.
Other similar governors use springs to provide the restoring force, and so do not need to be mounted vertically; they are still in use today to regulate the speed of certain electric motors and chiming clocks.

\begin{figure}[tp]
\centering
{\scriptsize (a)}\scalebox{0.63}{\begin{tikzpicture}
\tikzstyle{every node}=[font=\large\sffamily]
\pic {parallel-bkgd};
\pic {parallel={Controller\\\normalsize(arms)}{Plant\\\normalsize(incl Valve)}} ;

\node (r) at ($(start) + (0.3,0.3)$) {~~\textbf{r}\,=\,x$_{re}$} ;
\node (s) at ($(sum) + (1,0.3)$) {\textbf{s}\,=$\Delta$x} ;
\node (c) at ($(control) + (-2,0.3)$) {\textbf{c}\,=\,x} ;
\node (y) at ($(plant) + (2.5,0.3)$) {\textbf{y}\,=\,$\omega$ actual} ;

\end{tikzpicture}}\\\vspace{3mm}
{\scriptsize (b)}\scalebox{0.63}{\begin{tikzpicture}[font=\large]


\node[physfront] (p) at (3,-1) {$\strut\p_{c}$};
\node[absfront] (mp) at (3,3) {$\strut \omega_0, \omega_{re}$};  
\draw[->,double] (mp) -- node[right,near start] {$\IRc$} (p);

\node[absfront] (mprp) at (9,3) {$\strut s_1$};
\draw[absabsfront] (mp) -- node[above, near start,xshift=2mm] {\normalsize compute signal} (mprp);

\node[physfront] (ppr) at (9,-1) {$\strut\p'_c$};
\draw[physphysfront] (p) --  node[above] {\normalsize fly balls/linkages} (ppr);
\draw[->,double] (ppr) -- node[right] {$\Rc$} (mprp);


\node[physback] (p2) at (0.5,0.5) {$\strut\p_{p}$};
\node[absback] (mp2) at (0.5,4.5) {$\strut s_0, \omega_0$};
\draw[->,double] (p2) -- node[left] {$\RRE$} (mp2);
    
\node[absback] (mprp2) at (6.5,4.5) {$\strut s_1, \omega_0 + \delta \omega_0$};
\draw[absabsback] (mp2) -- node[above] {\normalsize restore speed} (mprp2);

\node[physback] (ppr2) at (6.5,0.5) {$\strut\p'_{p}$};
\draw[physphysback] (p2) -- node[above, near end] {\normalsize engine} (ppr2);
\draw[->,double] (mprp2) -- node[left,near end] {$\IRRE$} (ppr2);


\draw[physphysside] (p2) -- node[left, at end,xshift=-4mm]{$\boldsymbol{\omega}(t_0) =$\\speed} (p);
\draw[absabsside] (mp2) -- node[left, near end,xshift=-2mm]{$\omega(t_0)$} (mp);
\draw[physphysside] (ppr) -- node[right, near end,xshift=1.8mm]{$\textbf{s}(t_1) =$ steam valve} (ppr2);
\draw[absabsside] (mprp) -- node[right, near end,xshift=1mm] {$s(t_1)$} (mprp2);
	
\end{tikzpicture}}
\caption{\label{fig:guv_AR}
(a) The centrifugal governor (shaded) as part of the overall Control system.
See section~\ref{sec:guv eg} text for details.
(b) The centrifugal governor in  ART.
See section~\ref{sec:guv ART} text for details.
}
\end{figure}
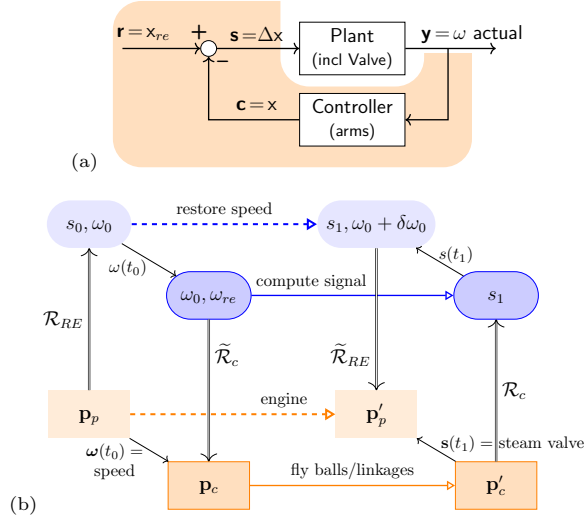

See figure~\ref{fig:guv_AR}a for the control system diagram.
The reference baseline $\mathsf{r = x_{re}}$ is a preset input to the controller (position of the collar); it is calibrated, and has to be changed to modify the preset value, by changing the linkage parameters $\mathsf{l_a}$ and $\mathsf{l_b}$.
The $\mathsf{y} = \omega$ output is sensed directly by the forces and linkages causing the fly balls to rise or lower, changing the position of the collar; $\mathsf{x}$ is the distance between the collar and the baseline.
The Controller is the governor mechanism, transducing angular velocity $\omega$ to position $\mathsf{x}$ through its physical properties of centrifugal force and linkage design.
The signal representing the difference between $\mathsf{x_0}$ and $\mathsf{x}$, $\mathsf{s} = \Delta \mathsf{x}$, is directly implemented by the lever attached to the valve.

\subsubsection{The centrifugal governor in ART}\label{sec:guv ART}

See figure~\ref{fig:guv_AR}b for how the centrifugal governor can be modelled in ART.

The components of the control system in figure~\ref{fig:guv_AR}a become the physical base layer of the ART cube in figure~\ref{fig:guv_AR}b.
The current engine speed $\boldsymbol{\omega}(t_0)$ is physically sensed by the mechanical system it is connected to;
along with the instantiation of the set value $\mathbf{x}_{re}$ through the linkage design,
this forms its current physical state $\mathbf{p}_c$.
The reaction of the fly ball/linkage system has the effect of raising or lowering the collar,
moving it end relative to the set point value,
forming its final physical state $\mathbf{p}_c$.
This movement directly signals how much the steam valve needs to be opened or closed through the connecting lever.

The abstract top layer of the cube
mirrors the electro-mechanical controller case, figure~\ref{fig:bimetallic_control}b,
with control variable temperature $T$ replaced by angular speed $\omega$,
and the signal being the new valve open value.

There is computation occurring in the centrifugal governor evolution in figure~\ref{fig:guv_AR}b,
corresponding to the 
calculation of the control system values, collar position $c = \mathsf{x}$ and steam valve signal $\mathsf{s = \Delta x}$, from figure~\ref{fig:guv_AR}a.
There is representation occurring,
as the speed $\omega, \omega_0$ and final steam valve state $s$ variables
in the abstract part of the compute cycle are
abstract representations of physical collar position and steam valve position in the physical controller.
Due to the mechanical setup, the $\mathbf{\Delta x}$ signal of the physical collar position is directly linked to the steam valve state.

In such a purely mechanical control system, 
again the transduction of the input (speed) and output (valve position) happens automatically due to the designed physical properties of the controller.
There is natural hysteresis and damping in the system:
the fly balls do not react instantaneously to a change in rotational speed,
and the speed does not react instantaneously to the valve being opened or closed more.

\subsection{A human-in-the loop control system}\label{sec:car}
In a human-in-the-loop control system, 
the controller is some human being manually adjusting the behaviour of the controlled plant.
\subsubsection{Example: a car heating system}\label{sec:car eg}

Modern car heating systems, like home heating controls (section~\ref{sec:dgital control}),
allow a given temperature to be set,
and then regulate the heater output to achieve this temperature.
Older car systems (like older house heaters) simply allow the power output to be set.
The driver or passenger has to adjust this directly to control the temperature.
This operator is the human-in-the-loop Controller of this open loop control system,
continually fiddling with the heating knob to achieve the desired temperature.

\begin{figure}[tp]
\centering
{
\scalebox{0.7}{\begin{tikzpicture}
\tikzstyle{every node}=[font=\large\sffamily]
\pic {serial-bkgd};
\pic {serial={Controller\\\normalsize(human)}{Plant\\\normalsize(heater)}} ;

\node (r) at ($(start) + (0.3,0.3)$) {~~\textbf{r}\,=T$_{re}$} ;
\node (e) at ($(sum) + (1,0.3)$) {\textbf{e}\,=$\Delta$T} ;
\node (s) at ($(control) + (2,0.3)$) {\textbf{s}\,=$\phi$} ;
\node (y) at ($(plant) + (2.5,0.3)$) {\textbf{y}\,=T} ;

\end{tikzpicture}}
}
\caption{\label{fig:open_loop}
The open loop car heater Controller (shaded) as part of the overall Control system.
See section~\ref{sec:car eg} text for details.
}
\end{figure}
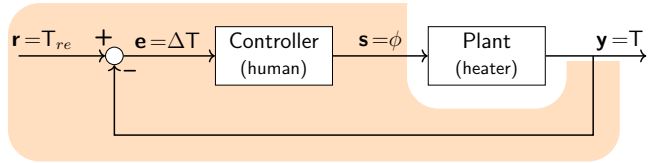

See figure~\ref{fig:open_loop} for the control system diagram.
The reference baseline $\mathsf{r = T_{re}}$ is the preset desired temperature (preset in this case to the desired temperature for the car cabin).
The $\mathsf{y = T}$ output (the current cabin temperature) is sensed by the Controller, in this case the human driver.
The difference between $\mathsf{T_0}$ and $\mathsf{T_{re}}$ is computed as the error $\mathsf{e} = \Delta\mathsf{T}$  (`too hot!', or `too cold!'),
and the controller calculates (the driver decides) the needed value of the signal $\mathsf{s = \phi}$, or how much the heating knob needs to be turned to get to the desired temperature.

\subsubsection{The car heating system in ART}\label{sec:car ART}

\begin{figure}[tp]
\centering
 \scalebox{0.7}{\begin{tikzpicture}[font=\large]


\node[physfront] (p) at (3,-1) {$\strut\p_{c}$};
\node[absfront] (mp) at (3,3) {$\strut T_0, T_{re}$};  
\draw[->,double] (mp) -- node[right,near start] {$\IRc$} (p);

\node[absfront] (mprp) at (9,3) {$\strut \phi_1$};
\draw[absabsfront] (mp) -- node[above, near start,xshift=2mm] {\normalsize compute signal} (mprp);

\node[physfront] (ppr) at (9,-1) {$\strut\p'_c$};
\draw[physphysfront] (p) --  node[above] {\normalsize human controller} (ppr);
\draw[->,double] (ppr) -- node[right] {$\Rc$} (mprp);


\node[physback] (p2) at (0.5,0.5) {$\strut\p_{p}$};
\node[absback] (mp2) at (0.5,4.5) {$\strut \phi_0, T_0$};
\draw[->,double] (p2) -- node[left] {$\RRE$} (mp2);
    
\node[absback] (mprp2) at (6.5,4.5) {$\strut \phi_1, T_0 + \delta T_0$};
\draw[absabsback] (mp2) -- node[above] {\normalsize restore temp} (mprp2);

\node[physback] (ppr2) at (6.5,0.5) {$\strut\p'_{p}$};
\draw[physphysback] (p2) -- node[above, near end] {\normalsize heater} (ppr2);
\draw[->,double] (mprp2) -- node[left,near end] {$\IRRE$} (ppr2);


\draw[physphysside] (p2) -- node[left, at end,xshift=-4mm]{$\textbf{T}(t_0) =$\\room temp} (p);
\draw[absabsside] (mp2) -- node[left, near end,xshift=-2mm]{$T(t_0)$} (mp);
\draw[physphysside] (ppr) -- node[right, near end,xshift=1.8mm]{$\boldsymbol{\varphi}(t_1) =$ knob angle} (ppr2);
\draw[absabsside] (mprp) -- node[right, near end,xshift=1mm] {$\phi(t_1)$} (mprp2);
	
\end{tikzpicture}}
\caption{\label{fig:car_ART}
The car heater open-loop control system in  ART.
See section~\ref{sec:car ART} text for details.
}
\end{figure}
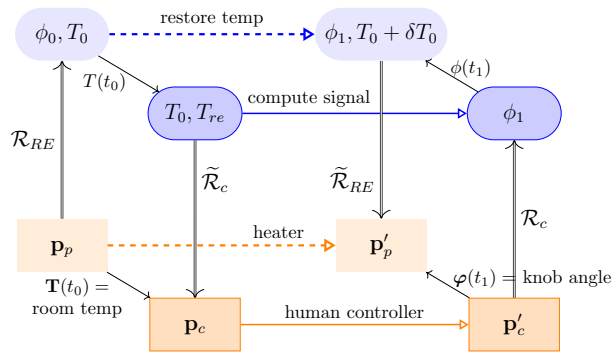

See figure~\ref{fig:car_ART} 
for how this digital control system can be modelled in ART.

The components of the control system in figure~\ref{fig:open_loop} become the physical base layer of the ART cube in figure~\ref{fig:car_ART}.
The current car cabin temperature $\mathbf{T}(t_0)$ is physically sensed by the physiology of the human driver,
as part of its current physical state $\mathbf{p}_c$.
The driver calculates the difference between $\mathbf{T}_0$ and $\mathbf{T}_{re}$ (is it too cold, or too hot, with some indication of by how much), and the new heater knob angle $\boldsymbol{\phi}$ needed to reduce this difference, through physiological and cognitive actions.
The calculated knob angle is physically decoded by the driver turning the knob,
affecting the heat produced
 in the next time step.

In the abstract top layer of the cube,
which is our model of the physical systems operation,
the plant's desired action is to restore the car cabin temperature $T_0$ to the preset value $T_{re}$.
It outputs the current temperature $T_0$ for (abstract) computation, which uses the error $\Delta T_0 = T - T_{re}$, and receives back the computed signal of how to achieve this, $\phi_1(\Delta T_0)$.
Its final state is $T_1 = T_0 + \delta T_0$,
the temperature it managed to achieve using the previous signal;
this will not necessarily be the desired temperature $T_{re}$, but, in a well-designed control system, will be making progress towards it.

There is computation occurring in the human controller evolution in figure~\ref{fig:car_ART},
corresponding to the 
calculation of the control system values, error $\mathsf{e} = \Delta \mathsf{T}$ and signal $\mathsf{s}$, from figure~\ref{fig:open_loop}.
There is representation occurring,
as the temperature $T, T_0$ and angle $\phi$ variables
in the abstract part of the compute cycle are
abstract representations of physiological and cognitive states in the human controller.
There is natural hysteresis in the system:
the human operator needs to be sufficiently experienced to implement a control algorithm that does not result in oscillating temperatures.

We have previously identified the plant (here the heater) as the physical representational entity
(or at least as a proxy for the design engineer);
see section~\ref{sec:translate}.
Here we have a human in the loop playing the role of the physical \textit{computer} within the control system;
this human is \textit{not} the representational entity in ART terms.
In this particular example, it is true that the purpose of the plant is to keep the operator warm.
However, in many open-loop systems,
the human operator is there to control the plants behaviour for some external reason: to fly the aeroplane safely, to control the thickness of rolled steel, etc.
Notice that figure~\ref{fig:open_loop} is identical in structure to figure~\ref{fig:digital_heater}:
the only differences are the identity of the Controller (human versus digital) and the signal (an angle to turn a knob, versus a voltage to apply to the heaters power source).
This precisely demonstrates the role of the human-in-the-loop: as a control computer, not as a representational entity.
So the presence of a human as part of a control system does \textit{not} imply that they are the representational entity.

\subsection{A digital plant control system}\label{sec:apollo}
In all the above examples, the nature of the controller varies from digital, through electro-mechanical and mechanical, to human,
whereas the plant being controlled is (electro-)mechanical in nature.
However, the plant could itself be a digital system,
having its behaviour controlled by a separate controller.
\subsubsection{Example: the Apollo Guidance Computer memory parity system}\label{sec:apollo eg}

Consider the Apollo Guidance Computer (AGC) system,
and specifically, the memory parity checking control \cite[p.HW-78]{Delco-Electronics1971}.
Each cycle, one word is loaded from storage memory into the processors local memory. 
This process is unreliable,
and the loaded word might be corrupted.
A parity check is performed:
if the loaded word has negative parity,
no corruption is detected, and the AGC proceeds as normal.
But if the loaded word has positive parity, and error has occurred,
and the parity checker signals the AGC to reset to a previous known state.

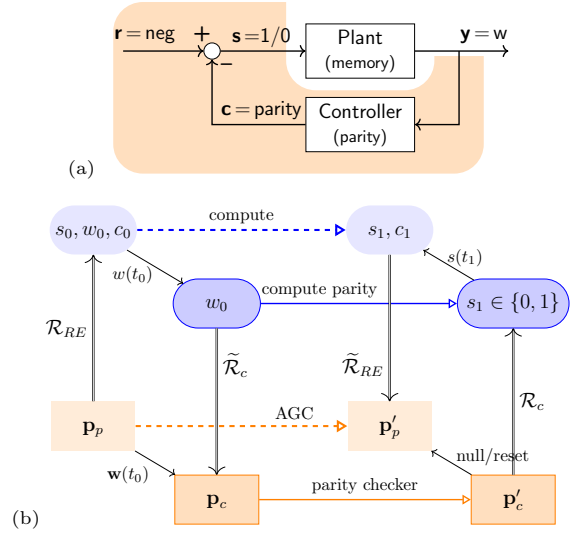
\begin{figure}[tp]
\centering
{\scriptsize(a)}\scalebox{0.65}{\begin{tikzpicture}
\tikzstyle{every node}=[font=\large\sffamily]
\pic {parallel-bkgd};
\pic {parallel={Controller\\\normalsize(parity)}{Plant\\\normalsize(memory)}} ;

\node (r) at ($(start) + (0.3,0.3)$) {~~\textbf{r}\,=\,neg} ;
\node (s) at ($(sum) + (1,0.3)$) {\textbf{s}\,=1/0} ;
\node (c) at ($(control) + (-2,0.3)$) {\textbf{c}\,=\,parity} ;
\node (y) at ($(plant) + (2.5,0.3)$) {\textbf{y}\,=\,w} ;

\end{tikzpicture}}\\\vspace{3mm}
{\scriptsize(b)}\scalebox{0.65}{\begin{tikzpicture}[font=\large]


\node[physfront] (p) at (3,-1) {$\strut\p_{c}$};
\node[absfront] (mp) at (3,3) {$\strut w_0$};  
\draw[->,double] (mp) -- node[right,near start] {$\IRc$} (p);

\node[absfront] (mprp) at (9,3) {$\strut s_1 \in \{0,1\}$};
\draw[absabsfront] (mp) -- node[above, near start,xshift=2mm] {\normalsize compute parity} (mprp);

\node[physfront] (ppr) at (9,-1) {$\strut\p'_c$};
\draw[physphysfront] (p) --  node[above] {\normalsize parity checker} (ppr);
\draw[->,double] (ppr) -- node[right] {$\Rc$} (mprp);


\node[physback] (p2) at (0.5,0.5) {$\strut\p_{p}$};
\node[absback] (mp2) at (0.5,4.5) {$\strut s_0, w_0, c_0$};
\draw[->,double] (p2) -- node[left] {$\RRE$} (mp2);
    
\node[absback] (mprp2) at (6.5,4.5) {$\strut s_1, c_1$};
\draw[absabsback] (mp2) -- node[above] {\normalsize compute} (mprp2);

\node[physback] (ppr2) at (6.5,0.5) {$\strut\p'_{p}$};
\draw[physphysback] (p2) -- node[above, near end] {\normalsize AGC} (ppr2);
\draw[->,double] (mprp2) -- node[left,near end] {$\IRRE$} (ppr2);


\draw[physphysside] (p2) -- node[left, at end,xshift=-4mm]{$\textbf{w}(t_0)$} (p);
\draw[absabsside] (mp2) -- node[left, near end,xshift=-2mm]{$w(t_0)$} (mp);
\draw[physphysside] (ppr) -- node[right, near end,xshift=1.8mm]{null/reset} (ppr2);
\draw[absabsside] (mprp) -- node[right, near end,xshift=1mm] {$s(t_1)$} (mprp2);
	
\end{tikzpicture}}
\caption{\label{fig:Apollo_CS}
(a) Apollo Guidance Computer controller (shaded) as part of the overall AGC system.
See section~\ref{sec:apollo eg} text for details.
(b) The Apollo Guidance Computer control system in  ART.
See section~\ref{sec:apollo ART} text for details.
}
\end{figure}

This is modelled as in figure~\ref{fig:Apollo_CS}a.
The word \textsf{w} is retrieved from memory (part of the Plant).
The Controller calculates the words parity,
and compares it to the reference parity (\textsf{r}, negative).
If the parity is negative, the signal is null (that is, just carry on);
if it is positive, the signal \textsf{s} is to reset the AGC to a previous known state, where all the words currently in local memory that have been pulled from storage have the correct negative parity.

\subsubsection{The Apollo Guidance Computer memory parity system in ART}\label{sec:apollo ART}
See figure~\ref{fig:Apollo_CS}b 
for how this digital control system can be modelled in ART.

The components of the control system in figure~\ref{fig:Apollo_CS}a become the physical base layer of the ART cube in figure~\ref{fig:Apollo_CS}b.
The loaded memory word $\mathbf{w}(t_0)$ is input to the parity checker;
along with the encoding of the set value $neg$,
this forms its current physical state $\mathbf{p}_c$.
The parity checker calculates the parity of the word,
forming its final physical state $\mathbf{p}_c$.
The result is either a null signal or a reset signal.

In the abstract top layer of the cube,
which is our model of the physical systems operation,
the plants desired action is to compute using correct memory values.
It outputs the current loaded word $w$ to the (abstract) controller, and receives back the computed signal of how to achieve this, corresponding to continue (no error detected), or reset (parity error detected).
Its final state is $c_1$,
the computational state it achieved while the parity checking was ongoing, or a reset to earlier computational state.
This approach allows the system to behave as if the memory loading occurs without error (albeit more slowly).

There is computation occurring in the parity checker.
There is representation occurring,
as the computational states 
in the abstract part of the compute cycle are
abstract representations of physically instantiated bit patterns in the digital controller.

\section{Discussion: Control Systems and Computing}\label{sec:discussion}

Here we have investigated the computational behaviour of control systems within the framework of Abstraction/Representation Theory. Using this, we have determined that all control systems -- simple or complicated; digital, mechanical, or human -- do indeed compute. The computation is performed by the Controller, at the behest of the Plant whose degrees of freedom it is controlling. There is no upper or lower bound given to the complexity of the computation being performed, and the existence of computing by the system does not in any way imply that they are universal, or Turing-equivalent, computers. 
They are, in fact, special purpose devices, often designed to perform just the computation needed for their specific task. In many cases this is quite simple, as indeed we can see from the extremely simple example systems used here. This was a deliberate choice, allowing us to demonstrate a general analysis of control systems, without danger of `smuggling in' computation within a complicated Controller. Since even these very basic control systems compute, more sophisticated cases certainly also do.

The implications of these results are broad, covering computing, control systems, debates around representation in science and engineering, and even cognitive theory. We have brought control system theory within the realm of computing theory. Control systems are \textit{computing}, but are not identical with \textit{computers}: most computers are not Controllers within control systems. Whether computing exists in a system under consideration can now help determine if it is a control system. More precisely, each of the ART components of section \ref{sec:ART-Control} need to be located in order for a control system to be identified. We have given five examples of such a process of identification and analysis, covering a range of technologies and implementations (section~\ref{sec:control_ART}), and many more instances and categories of control systems exist.


We now draw out the specific implications of our work along several axes. 
Firstly, we discuss to use of the centrifugal governor example in cognitive theory.
Secondly, we consider the roles played by the Representational Entity and any humans (and their cognition) within the system.
Thirdly, we examine the precise relationship between control systems and computers that has been revealed by the analysis here. 
Then we discuss the relations between the different parts of a control system, Plant, Controller, and computation, as they mediate representational activity between them in important ways. 
Finally, we see how historical control systems (such as the steam governor) allow us to identify novel computing devices, and the deeper understanding of control systems this also provides.

\subsection{The centrifugal governor in cognitive theory}\label{sec:vanG}

In section \ref{sec:control_ART} we used a variety of examples to demonstrate that control systems have the three qualities necessary for computing: (i) a representation from physical process to abstract computation; (ii) an $\epsilon$-commuting diagram, demonstrating that the physical process performs the desired computation in the context of a representation; (iii) a representational entity, which `owns' the representation, and by which the computation is desired. One of these examples is the mechanical centrifugal governor.

This governor has a rich history of use within the theory of cognition. While there is no argument that \textit{some} mechanical systems can compute \citep{Yasuda:2021:mechcomp}, the argument around the governor is whether that system in particular is computing, and what that means for whether the brain itself is a computer. We now apply our analysis of the governor specifically within that arena, showing that one common use of the governor example is not correct.

The debate the governor is used within is whether computationalism is a correct theory of the brain as a computer. An excellent summary of the state of play is \cite{Baltieri2020}. Simplified, computationalism argues that because the action of the brain appears (to some) to `look like' the action of a computer, that therefore the brain \textit{is} a computer. The governor is used as a counter-argument, that this is a device that also looks like it is performing computation, but it is not. Therefore, the counter-argument goes, computationalism is an invalid method of arguing that the brain is a computer.

An often-used version of this counter-argument is given by \citet{Van_Gelder1995}. While it is clear from our above analysis of the governor that this counter-argument fails since the governor is itself computing, it is also instructive to see exactly how the ART treatment differs in specific points of argument. These points typically revolve around concepts of dynamical systems and of representation, with the governor used as a simple example to help clarify the points raised.

\citet{Van_Gelder1995} argues that dynamical systems (including the  governor) are different from computational systems: ``Rather than computers, cognitive systems may be dynamical systems; rather than computation, cognitive processes may be state-space evolution within these very different kinds of systems.''
However, the state space evolution of dynamical systems can  be used to compute \citep{Stepney:2012:DynSys};
for example,
the very field of Neural Networks in general, and Reservoir Computing in particular \citep{Jaeger2001,Appeltant2011}, is based on dynamical systems models;
physical reservoir computing can be cast in terms of ART \citep{Stepney-2024-NACO}.

More pertinently from the point of view of ART, \citet{Van_Gelder1995} also argues that the governor is nonrepresentational: ``arm angle and engine speed are of course intimately related, but the relationship is not representational'', and ``mere correlation does not make something a representation''. However, we have argued above that there \textit{is} a representational relationship present: the abstract rotation speed $\omega$ is an abstract representation of the physical 
\textit{height of the collar} $\mathbf{x}$ (figure~\ref{fig:guv}b).  This height $\mathbf{x}$ is itself non-trivially related to arm angle (determined by the actual rotation speed), through the mechanism of the engineered linkages,
to establish an offset from a given set point.

In order to argue further against representation, and even correlation, \citet{Van_Gelder1995} notes that
``even as the arms are falling, more steam is entering the piston, and hence the device is already working; indeed, these are exactly the times when it is most crucial that the governor work effectively. Consequently, no simple correlation between arm angle and engine speed can be the basis of the operation of the governor.''
However, we note that the system is not instantaneous.
The arms have inertia: they take time to fall;
the collar takes time to move, to perform its computation;
the changed steam pressure takes time to change the speed of the shaft.
Granted, the system is in some sense continuous:
the cycle time of the computation may be better modelled as the $dt$ of calculus, not the $\delta t$ of discrete time systems.
However, the engineered system includes feedback delays as modelled in the ART compute cycle.

In order to help discover whether a relationship is representational,
\citet{Van_Gelder1995} presents a helpful criterion:
``A useful criterion of representation---a reliable way of telling whether a system contains them or not---is to ask whether there is any explanatory utility in describing the system in representational terms.''
\citet{Baltieri2020} express the criterion in more explicit terms: ``is it useful to explain a Watt governor in representational terms or, in other words, does taking the governor flyball arm angle to \textit{represent} the speed of the engine help us understand the workings of the governor?''

We can use this criterion, and in fact argue for the presence of such explanatory utility.
First, the representation helps explain the very design of the  governor:
the need for the rotation speed $\omega$ to be a representation of the height of the collar $\mathbf{x}$  
(not of the mere arm angle) explains the engineered linkage design.
Second, it allows us to unify the controllers in control systems,
whether explicitly computational, or mechanically engineered,
or human-in-the-loop,
under the single umbrella of a class of computational systems.

So, along with the analysis of section \ref{sec:control_ART}, we argue that the governor is indeed computing.


\subsection{Control systems, representational entities, and humans in the system}


In the context of ART, control systems, as engineered systems that necessarily contain \textit{computing}, also necessarily contain \textit{representation}. The identification of representation (and indeed of the whole abstract layer of control) is also a crucial part of identifying a control system. This necessitates the identification of the representational entity (RE) being used by the compute cycle 
\citep{Horsman2014,StepneyKendon2019}, responsible for the representation of the computation input, evolution, and output in abstract terms.
This RE is not usually co-extensive with the computing device. In classical computing, the RE is typically the human using the computer. However, the RE need not be human: it could, for example, be a bacterium performing intrinsic computing \citep{Horsman2017bio}. Thinking about physical computing through the lens of control theory helps sharpen, and broaden, the notion of RE. Here, we have determined that the RE in a control system is the controlled Plant. More precisely, the Plant is a proxy for the real RE: the engineer who designed it. 

These divisions between Controller, Plant, and RE are particularly important to keep clear in the case of control systems that include human beings. All control systems compute; in an open loop control system in general, the human operative is the computer, not the RE. There is something perhaps slightly counter-intuitive in thinking of the human Controller not being in control, and the Plant apparently `requesting' the computation. The request must generally be thought of coming ultimately from the engineer who designed the control system, and is responsible for the representation used by the computer. 

There are human-in-the-loop control systems, however, where the RE and the Controller are same person. The example given in  section~\ref{sec:car} of the car heating system and driver is one such case. The human driver is setting the parameters that they want the Plant to operate within (that the inside of the car remains at a pleasant temperature), and then placing themselves within the system as Controller, performing computation in order to maintain the Plant within those parameters. The human has in that case engineered a control system comprising the inside temperature, the heating, and themselves. 

These types of control systems are an example of `human as computer', where humans perform computational tasks. This class includes examples ranging from performing simple mental arithmetic, through to the original human `computers' and today's crowd-sourcing tasks \citep[sec.9]{Horsman2014}. In the example of a human adjusting car air temperature, the RE is the same entity as the computer, but this need not be the case for control systems in general, as with other human computation (for example, consider a car driver adjusting the temperature to suit the stated comfort of a passenger, who then becomes the RE).

In addition to providing the tools that define humans as REs vs Controller or Plant,  we have also been able to make a contribution to the debate around whether the human brain itself is a computer. The  case of the (mechanical) governor has been used to argue against computationalism, as an example of a system that (supposedly) can appear computing-like but is not (section \ref{sec:vanG}).
We have demonstrated that the governor is indeed computing (section \ref{sec:guv}); hence the discipline of cognitive philosophy should choose a different counterexample for their arguments. Additionally, we have argued that the human in the open-loop control system is also computing, in the same abstract if not physical manner as the governor. Therefore, the arguments about whether cognition, or mechanical controllers, are computing requires careful definition of (physical) computing and identification of representational activities.


\subsection{Control systems vs computers in ART}


The way we have identified computation within control systems is not simply by identifying the \textit{existence} of representation (and associated RE), but depends critically on the \textit{way} that that representation is processed. Computing is a specific cycle in ART: the use of a physical system (the computer) to predict the result of an abstract evolution (the result of the computation). In section \ref{sec:ctrlsys}, we identified computing in control systems by identifying the compute cycle as necessarily present in every control system. 

We have shown that all control systems contain computation (and hence a computer) as part of their operation, 
but we have \textit{not} shown that a control system \textit{is} (necessarily) a computer. This statement may appear paradoxical;  it is important for a full understanding of our results to see how it is not. Control systems in our analysis remain fully distinct from computers. While a computer can be a control system (for example, the AGC, section~\ref{sec:apollo}), not all control systems are computers, and not all computers are control systems. These distinctions can be seen by considering two key elements of a control system: the level (abstract vs physical) that a control system operates at, and the specific computation that is performed within such a system.

\subsubsection{Abstract vs physical outputs}

Figure \ref{fig:ART_ctrl} shows how the control system cycle in ART contains, embedded within it, a compute cycle. However, the complete control system cycle is not identical with a compute cycle. It forms a distinct representational cycle (much as science and engineering are distinct in ART \citep{Horsman2014}); the presentation of this separate control system cycle is one of the results of this paper.

In the compute cycle (figure~\ref{fig:repn-b}), abstract quantities (of the problem) are encoded in physical states (of the computer), processed through physical evolution, then represented back as abstract quantities (the solution). Crucially, for a compute cycle, the inputs and outputs are abstract. A computer is a physical device for processing abstract data and gaining abstract outputs. 

By contrast, most of the control system cycle (figure~\ref{fig:ART_ctrl}) concerns the engineering and operation of a physical system. When the control system runs, it is the physical evolution of the Plant, $\mathbf{H}_{p}$, that begins. When the Controller comes into play, we have the compute cycle as a sub-cycle of the control system cycle. The relevant physical quantities of the Plant are abstractly encoded in the state of the Controller, and that abstract encoding then processed physically by the Controller, before being used to change the state of the Plant. This last step is key: the abstract result of the Controller (the signal $s(t_1)$) is decoded back to a physical state of the Plant. An interruption of the control system cycle at the abstract level leaves an incomplete cycle; only when these data are used to change the control system \textit{physically} does the cycle completes.

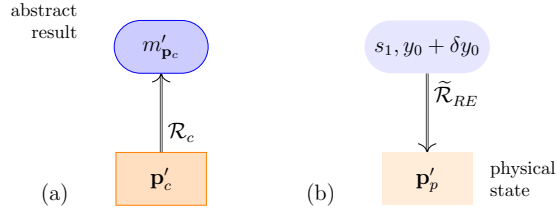
\begin{figure}[tp]
\centering
 \scalebox{0.7}{\begin{tikzpicture}[font=\large]

\node[absfront] (mprp) at (3,1.5) {$m_{\p_c}^\prime$};
\node[physfront] (ppr) at (3,-1) {$\strut\p^\prime_c$};
\draw[->,double] (ppr) -- node[right, near start] {$\Rc$} (mprp);
\draw[align=right,font=\normalsize] (0.8,2) node {abstract\\result};
\draw (1,-1.3) node {(a)};

\node[physback] (p) at (8,-1) {$\strut\p_p^\prime$};
\node[absback] (mp) at (8,1.5) {$\strut s_1, y_0 + \delta y_0$};
\draw[->,double] (mp) -- node[right, near start] {$\IRRE$} (p);
\draw[align=left,font=\normalsize] (9.8,-1) node {physical\\state};
\draw (6,-1.3) node {(b)};
	
\end{tikzpicture}}
\caption{\label{fig:difference}
Difference in outputs for (a) compute cycle and (b) control
system cycle. For a compute cycle the output is the
\textit{abstract} result of the \textit{computation}. For a control system
cycle it is the final \textit{physical} state of the \textit{plant} (representational entity).
}
\end{figure}

Figure \ref{fig:difference} shows this difference between the final steps that produce the outputs of the (a) compute  and (b) control system  cycles. The are, in effect, mirror images of each other, with the role of abstract and physical reversed. For a computer, the final physical state of the computer, $p^\prime_c$, is converted via the representation relation $\Rc$ into the abstract output $m_{\p_c}^\prime$. By comparison, the final step in the control system cycle is when the abstract outcome of the Controller's computation of the $s_1(\Delta y_0)$ (where $\Delta y_0 = y_0 - y_{re}$) is converted, via the reversed representation relation $\IRRE$, into the \textit{physical} output of the plant $p^\prime_p$. 

It is therefore the \textit{physical} state of the control system that is its output. A control system in general is much more like a tool such as an axe or a bridge than a computer: in order to describe what this system is and how it functions, $m^\prime_{\p_c}$ is not relevant. Simply put, it is not necessary to identify an abstract layer for the control system considered as a black box in order to explain what it is doing. For computers, by contrast, what is going on needs to be understood in terms of abstract dynamics \textit{for the system as a whole}. 

Two of our examples highlight this contrast in particular. Firstly, and perhaps counter\-intu\-itively,  consider the Apollo Guidance Computer example of section~\ref{sec:apollo}. In this case we have a computer (the AGC), which contains a control system as part of its physical operation (the memory retrieval error checking system in the main processor), which in turn contains a computation as part of the Controller (the calculation of the parity of the word pulled from long-term storage into the processor). 
In this case, we have a control system that is itself a computer (the AGC). This is possible because, while computation is abstract, \textit{computers} are physical systems. Here this physical system also a control system, because of the hardware error correction. As we have demonstrated, this necessitates that computation is happening as part of the control system functioning. However, this is \textit{not} the same as the computation that is being done by the control system as a whole when regarded as a computer. The AGC sent people to the moon. The memory retrieval control system computed the parity of a bit string. These are computations that differ both in degree and in kind (more on this below), and are independent. 

From the point of view of the error correction control system, including its parity computation, it does not matter that this physical system is computing how to get people to the moon, it is simply a physical system that is being controlled subject to the dynamical constraints given by the RE (which does care what it is doing). To view the AGC as a computer rather than a physical control system requires a representation of the AGC as a whole, and an abstract evolution that the control system is being used to predict: a representation entirely separate from the representation of parity within the Controller of the error correction system. In the absence of this larger representation, a control system in itself is (simply) a physical control system.

The mechanical governor of section \ref{sec:guv} is one such instance of a physical control system that is not itself a computer. The desired outcome of the engineered system as a whole is a physical device that regulates the flow of steam through a physical system. This outcome is the $\p_p^\prime$ of figure \ref{fig:guv_AR}b. That physical layer is all that is needed to understand the operation of the steam governor in its entirety. Unlike with a computer, there is no need to construct an abstract layer of information processing for the steam governor as a black box to explain what it does.

This last point may help explain why cognitive theory has gone wrong using the governor in its arguments on the computational nature (or not) of the brain. The argument there essentially boils down to: the  governor (i) cannot be computing (despite perhaps looking like it does) because (ii) it is obviously not a computer; hence, while the brain may look like it is computing, this does not necessarily mean that it is. We can see now from the ART analysis that the second point is correct, but the first is not. The governor, considered as a whole, is not a computer. It does not perform compute cycles but rather control system cycles. It does, however, \textit{contain} computation (therefore compute cycles) in the action of the Controller. 

The representation relation used there ($\mathcal{R}_c$ in figure \ref{fig:guv_AR}b) does not give a representation for the steam governor as a whole (Plant + Controller) to be acting as a computer; and neither does the functional understanding of the operation of a centrifugal governor within a steam engine require one. What matters there is that the steam governor governs the flow of steam through the engine, not that it predicts the outcome of a desired abstract evolution. We see again how ART enables us to make these distinctions that can otherwise be lost, in particular in areas that have historically lacked such precision tools. It is very likely that ART would be of further utility in cognitive theory, especially around questions of computation and representation, and we offer it accordingly.

\subsubsection{Feedback: computing the $\Delta y$}

This distinction between control system and compute cycles allows us to conclude that control systems are not necessarily computers: while they \textit{can} be computers, as with the AGC, not all control systems are computers. This leaves us with the inverse question: might all computers be in fact control systems; is the category `computer' a sub-type of the broader class `control system'?

To see that this is not the case, consider how we might go about fitting the ART compute cycle into the model of the control system cycle. A control system is characterised by repeated, chained, uses of the control system cycle, as seen in figure~\ref{fig:unwound}. This is not typically the action of a computer, which tends to execute a set of instructions in order to go from the initial to final states. It is the steady state of the control system that is generally required, not a fixed end point. Nevertheless, for the sake of argument, one might try to fit a model of this general computing into a control system model by considering each implementation of a command as akin to one use of a control system cycle. In each case, the physical device has been evolving under its natural dynamics since the last use of the cycle, in a computer since the last command was given, and in a control system since the last action of the Controller. Then an external action changes this dynamics, by a command (computer) or action of the Controller (control system). And so the system repeats. 

So far so similar. The difference lies in the nature of the external action that changes the state of the system. In a generalised computing scenario this is the next in a pre-determined sequence of operations that does not, in general, depend on the input it has been given at this particular step.
By contrast, in a control system the action that changes the state is given by the operation of the Controller that specifically computes the difference $\Delta y$ between the actual state of the Plant, $y$, and the desired state of the Plant, $y_{re}$. In order for the Controller to operate, the desired state $y_{re}$ \textit{must be known in advance} and encoded, also in advance, into the construction of the Controller. 

The changes of state in a control system are therefore necessarily the result of feedback, measured against a known, ideal state. 
This contrasts sharply with generalised computing, where the whole point of the use of a computer is to predict the outcome of an abstract evolution that is \textit{not} known in advance. In general, one does not build and run a computer to compute an outcome that is already known.

Again the AGC clearly shows this difference between control and computing. We have a very specific type of computing -- hardware error correction -- that is also a control system. Error correction is a specialised kind of computation where one already `knows the answer': a computation is performed comparing the actual state (in our AGC example the parity of the retrieved word) with the desired, pre-known outcome (negative parity). Error correction is a routine run in addition to the \textit{actual} computation being performed by a processor, with the aim of catching errors in that actual computation. The actual computation itself cannot be checked directly for errors precisely because its correct, ideal state is \textit{not} known in advance. 

We can therefore see that control systems as a class are entirely distinct from computers. While all control systems contain computation, and some may in fact be computers (as with the AGC), control systems are not necessarily computers (because of the difference in the abstract/physical types of input/outputs), and computers are not necessarily control systems (because of the requirement for continuous feedback against a known state).


\subsection{Computing the $\Delta y$ in control systems}


We have shown that a Controller in a control system is necessarily computing. It may seem tempting to conclude that it is therefore somehow the most important, load-bearing part of the control system. However, our analysis shows that this conclusion would be false. To see why, we need to consider the relations between the Plant, the Controller, and the computation performed.

As we have seen, the action of the Controller is not a general-purpose computation but the same form of special-purpose computation happening across widely different systems and computational substrates, namely, the computation of the difference $\Delta y$ between the desired state of the control system Plant, $y_{re}$ and its actual state $y$ at time $t$. This $\Delta y$ is then translated back by the Controller into a control signal (feedback) to change the state of the Plant. 
This interrelation of the action of the Plant and the computation performed by the Controller is a defining element of control systems. 
The action of the control system is not dominated by the action of one or other of the Plant or Controller, but requires \textit{both} to be acting together in a complex feedback loop to reach the desired outcome. 

The Plant evolves under its own internal dynamics in between feedback cycles from the Controller, which computes. The Plant dynamics, giving the $\delta y$, are in no way subordinate to the computation of the Controller. For example, in the digital thermostat heating system of section~\ref{sec:dgital control} the final outcome -- a comfortably-heated house -- cannot be achieved simply by an electronic thermostatic circuit. The Plant produces what most users would consider to be the most important part of a heating system, namely the \textit{heat}.

Control systems are in this sense not combined systems of Plant+Controller, but truly \textit{heterotic} systems~\citep{Horsman2015}. In a heterotic system, the action of the system as a whole is much more than a simple addition of the actions of its constituent parts. In some heterotic computing system the combination of two computational systems can compute problems in a higher complexity class from those of the constituent systems individually. In some cases, the addition of only a very simple extra system is sufficient to produce this effect \citep{Anders2009}.

We see an analogous heteroticity in action in control systems, where the Controller is often performing only a very simple computation and action, compared to the complex system dynamics of the Plant. In our three examples of heating systems (section~\ref{sec:control_ART}), it is arguably only the third example, where a human acts as Controller, that the complexity of the Control computation and action approaches (or surpasses) the complexity of the dynamics of the Plant, namely the system producing the heat in the first place. 

We are not, in general, talking about computational complexity as regards the Plant and the control system as a whole (as control systems are not necessarily computers). However, one can reasonably talk about the complexity of dynamics of a physical system, and certainly about the complexity of its engineering. 

It is clear from our different examples, and the ART analysis presented here, that a key use-case of control systems is where the bulk of the work done by the control system is performed by the complex dynamical evolution of the Plant. However, Plant dynamics alone are not enough to engineer the desired system. Without the Controller operations, the Plant will continue evolving under $\mathbf{H}_\p$; for example, an unregulated heating system would continue to heat the environment, and would quickly become dangerous. The addition of a (possibly very) simple dynamical and computational Controller (such as a thermostat) renders the whole system far more dynamically complex than the individual components, and of much greater engineered utility. The Plant is still doing the bulk of the work, the Controller is often very simple, but the combination in a control system cycle of continuous feedback and monitoring allows the (often) far more complex system to be produced, which could otherwise be technologically unobtainable.

The relative importance and complexity of Plant and Controller can be seen very clearly in the final example of the Apollo Guidance Computer (sec.\ref{sec:apollo}). In this situation, the Plant, the Controller, and the control system itself are all, individually, computers (that is, they each perform full compute cycles when considered in isolation and in totality). The Plant (the main processor) is the main bulk of the AGC. The Controller performs the very limited computation of calculating the parity of each word retrieved from storage, and comparing it against the desired outcome (negative parity). 
This parity checking is a simple {\sc xor} operation plus restart. {\sc xor} on its own is a simple gate, and is not Turing complete. The Controller, therefore, is capable only of a very restricted set of computation (in fact, parity checking). By contrast, the Plant in the AGC is Turing complete,  performing far greater computations than those of the parity-checking Controller. The AGC as a whole (Plant plus parity error correction) is also Turing complete, and got people to and from the moon. The bulk of the `computational heft' of the AGC comes from the \textit{Plant}, rather than the `primitive' Controller. The combination of the two in a control system enabled the AGC as a whole to be used in moon missions.

The AGC example, perhaps more than any of the others, shows us just how much we should avoid conceptually prioritising the Controller simply because we have identified that it is computing. Control systems are not made up of unthinking, plodding Plants with uninteresting dynamics, and smart, resourceful Controllers. 
Plants are the users of the computer that is the Controller, and Plant dynamics $\delta y$ perform the bulk of the work in control systems, enabling the simple computation of the $\Delta y$ to lead to complex engineered systems. Of course, as our analysis also shows, neither should we discount the contribution of the Controller, which performs the vital control computation,  in systems that in general are not otherwise computing. Control systems are a true heterotic union of the two, Plant and Controller, managing representational activity between themselves in order to create highly complex engineered physical and computational systems.



\subsection{Identifying computing in historical control systems}


The identification of computing happening in all control systems has, perhaps surprisingly, implications for our understanding of computers and computer science. We are now in a position to identify a much broader class of computing devices than has previously been realised. In particular, as we see with the mechanical governor, this includes historical devices from systems (such as steam engines) that have been considered entirely distinct from any computing devices at that time. 

It is uncontroversial to say that devices such as the Babbage engine \citep{Swade1995,Fuegi2003} or the Pascaline \citep{Chapman1942,Gradstein1962} compute. We can now include in this set control system devices such as our simple mechanical  governor from the dawn of the industrial revolution. This helps deepen our understanding of what computing is and how it is implemented, especially when such newly-discovered computers use mechanisms and substrates that are considered `unconventional' \citep{Stepney2018}. The governor is not classical Turing-style computing; it is analogue, includes constant feedback loops, and does not give an abstract output to a human user: the Plant is the user. And yet it computes.

How have we been able to identify these new computers, in devices that are not themselves new (or newly discovered)? 
Firstly, this is because of the ability of ART to analyse control systems in terms of representational activity within and between their component parts. ART also gives a clear structure and set of criteria that computing must satisfy, which significantly helps in identifying computation.

Secondly, it is because of the highly specific, and in fact quite unusual, nature of the computation being performed in control systems, that these historical computers can now be identified. 
The computation is, crucially, in general \textit{not performed for the user of the plant}. The computation of a simple $\Delta y$ is often of no interest as a computation to an external user. Unlike, say, the Babbage engine, an external user of a centrifugal governor generally has no interest in the result of the computation performed by the Controller; the user requires a functional automatic steam regulator, not an abstract computational result. So a steam governor is not used as a computer \textit{by the external user}, but only internally in the system. Indeed, not only is the computation of the Controller not performed for an external user, it is generally not even output to them. They can infer the result of the computation only through the action (output) of the system as a whole.

Finally, it is likely the simplicity of the Controller's computation that explains why it has remained hitherto unrecognised.
We discuss above how computing the $\Delta y$ can be a basic, even primitive, computation. This allows it to be performed on devices that are technologically much simpler than modern computers. The Controller in general, especially in historical systems, is nowhere near complex enough to compute the outcome for the evolution of the control system as a whole. A simple Controller, without access to the Plant's native evolution, generally does not have the degrees of freedom to represent the desired final state of the Plant, let alone compute what it should be. 
Computing a much simpler $\Delta y$, however, when combined with the $\delta y$ of the Plant, enables the final system to be produced. A complex evolution can therefore be obtained in an era where the computational simulation of this evolution is technologically impossible.

This final point leads us to a much deeper understanding of how control systems work as technology. Engineering and computation are fused to create systems that can be capable of much more complex behaviour than either technology can produce alone. Control systems emerge as a powerful tool for bootstrapping technological -- including computational -- ingenuity.


\section{Conclusions}


ART has enabled us to show in detail how representation is managed within control systems, what role is played by the Representational Entity, and the relationship and separation between the classes of computers and control systems. We have seen how the computing Controller is not the dominant part of the control system, but works together with the inherent dynamics of the Plant in a way that is capable of significantly increasing the technological (including computational) complexity of the joint system. 

We have been able to achieve the insights presented here because of the power of the ART analysis when applied to control systems. This has allowed us to answer not just \textit{when} a control system is computing (they all are), but \textit{which part} is computing (the Controller), at \textit{whose request} (the Plant, the Representational Entity for the computation), and furthermore \textit{what} it is computing (the $\Delta y$ between the desired and actual Plant evolutions).

Our results have implications for areas as diverse as the computational abilities of the human brain, and the discovery of novel historical computers.
It is likely that our results will have a broader role to play in the future in debate on the use of representation within engineered technological systems, as well as furthering the work of defining when, how, and in what particular way some of those systems compute.

\subsection*{Acknowledgements}

VK acknowledges funding by UK Engineering and Physical Sciences Research Council
(EPSRC) grant EP/L022303/1,
and 
by ARIA `Nature Computes Better' grant 20240730 NACB-SE02-P02.

\bibliographystyle{unsrtnat}  
\bibliography{guv}

\end{document}